\documentclass[manuscript]{acmart}

\AtBeginDocument{%
  }

\setcopyright{none} 
\renewcommand\footnotetextcopyrightpermission[1]{} 

\usepackage{subcaption}
\usepackage{fvextra}
\usepackage[title]{appendix}
\usepackage{multirow}
\usepackage{multicol}
\usepackage{booktabs}
\usepackage{makecell}

\usepackage{enumitem}
\usepackage{ltxtable}
\usepackage{listings}
\usepackage{comment}
\usepackage{fancyvrb}
\usepackage{caption}
\usepackage{cprotect}
\usepackage{fontawesome5}
\usepackage{bm}
\usepackage{tikz}
\usepackage{algorithm}
\usepackage{algorithmic}
\usepackage{xspace}

\newif\ifshowcomments
\showcommentsfalse
\ifshowcomments
    \newcommand{\jz}[1]{{\color{blue}[JZ: #1]}}
    \newcommand{\xt}[1]{{\color{pink}[Xiaoting: #1]}}
    \newcommand{\yz}[1]{\textbf{\textcolor{teal}{[yingzhe:\ #1]}}}
\else
   \newcommand{\jz}[1]{}
   \newcommand{\xt}[1]{}
   \newcommand{\yz}[1]{}
\fi
\newcommand{\circlednum}[1]{%
  \tikz[baseline=(char.base)]{
    \node[shape=circle,fill=black,text=white,inner sep=0.01pt] (char) {#1};}}
\newcommand{\eg}{\emph{e.g.}\xspace}

\begin{document}

\title{Navigating the Unknown: A Chat-Based Collaborative Interface for Personalized Exploratory Tasks}


\author{Yingzhe Peng}
\authornote{Work is done during an internship at Microsoft.}
\email{yingzhe.peng@seu.edu.cn}
\affiliation{%
  \institution{Southeast University}
    \country{China}
}

\author{Xiaoting Qin}
\email{xiaotingqin@microsoft.com}
\affiliation{%
  \institution{Microsoft}
      \country{China}
}

\author{Zhiyang Zhang}
\affiliation{%
  \institution{State Key Laboratory for Novel Software Technology, Nanjing University}
    \country{China}
}

\author{Jue Zhang}
\authornote{Corresponding author.}
\email{jue.zhang@microsoft.com}
\affiliation{%
  \institution{Microsoft}
    \country{China}
}

\author{Qingwei Lin}
\affiliation{%
  \institution{Microsoft}
      \country{China}
}

\author{Xu Yang}
\affiliation{%
  \institution{Southeast University}
      \country{China}
}

\author{Dongmei Zhang}
\affiliation{%
  \institution{Microsoft}
      \country{China}
}

\author{Saravan Rajmohan}
\affiliation{%
  \institution{Microsoft}
      \country{USA}
}

\author{Qi Zhang}
\affiliation{%
  \institution{Microsoft}
      \country{China}
}

\renewcommand{\shortauthors}{Peng et al.}

\begin{abstract}



The rise of large language models (LLMs) has revolutionized user interactions with knowledge-based systems, enabling chatbots to synthesize vast amounts of information and assist with complex, exploratory tasks. However, LLM-based chatbots often struggle to provide personalized support, particularly when users start with vague queries or lack sufficient contextual information. This paper introduces the \underline{\textbf{C}}ollaborative \underline{\textbf{A}}ssistant for Pe\underline{\textbf{r}}sonalized \underline{\textbf{E}}xploration \textbf{(CARE)}, a system designed to enhance personalization in exploratory tasks by combining a multi-agent LLM framework with a structured user interface. CARE's interface consists of a Chat Panel, Solution Panel, and Needs Panel, enabling iterative query refinement and dynamic solution generation. The multi-agent framework collaborates to identify both explicit and implicit user needs, delivering tailored, actionable solutions. In a within-subject user study with 22 participants, CARE was consistently preferred over a baseline LLM chatbot, with users praising its ability to reduce cognitive load, inspire creativity, and provide more tailored solutions. Our findings highlight CARE’s potential to transform LLM-based systems from passive information retrievers to proactive partners in personalized problem-solving and exploration.

\end{abstract}


\maketitle

\begin{figure}[!htb]
	\centering
	\includegraphics[width=1\textwidth]{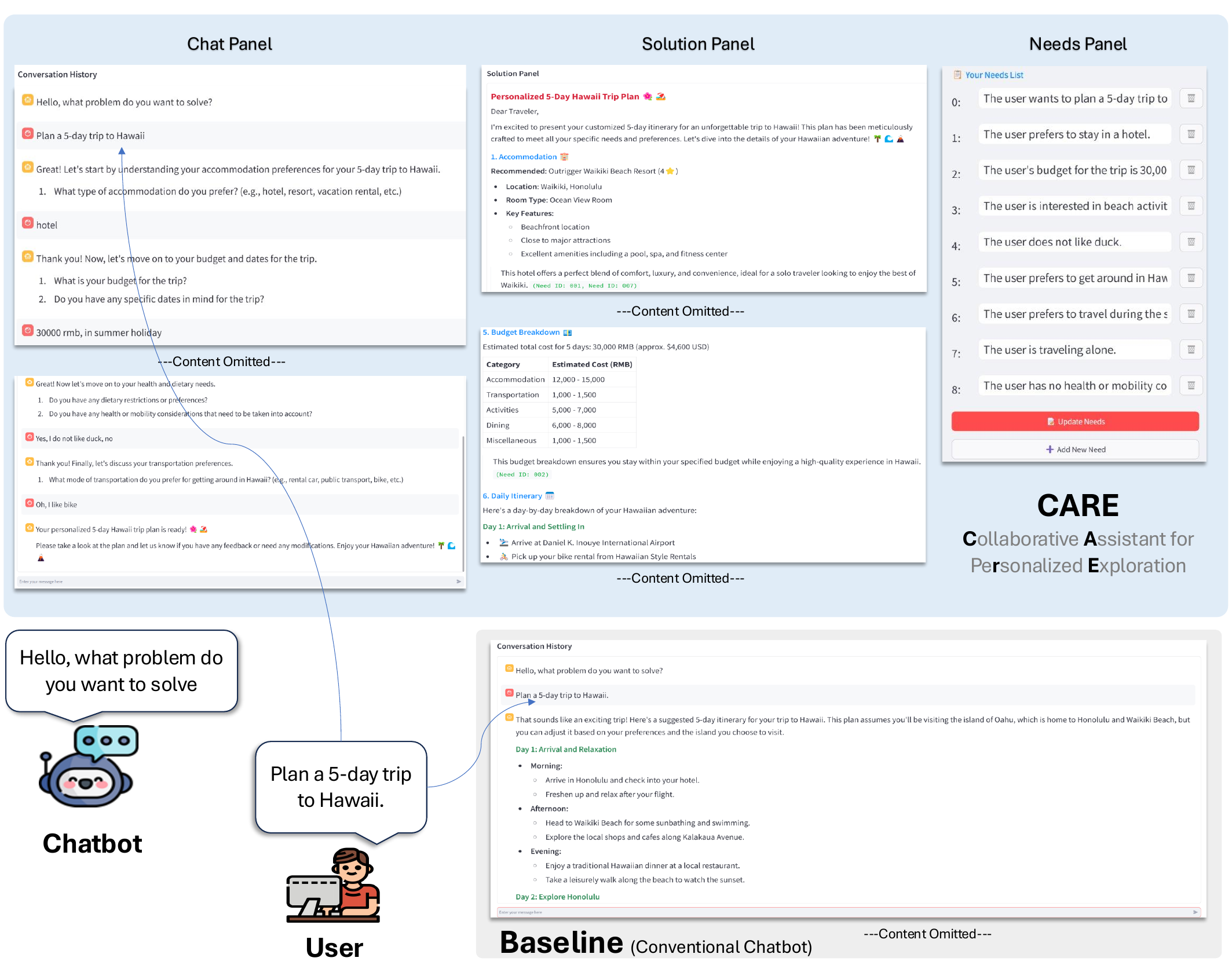}
	\caption[Figure]{Comparison of the UI and interaction styles between the CARE System and Baseline System. At the top is the CARE System, displaying the Conversation Panel, Solution Panel, and Needs Panel. The CARE System actively prompts the user, gathering their needs before creating a tailored plan. In contrast, the Baseline System, shown at the bottom right, features only a Chat Panel and tends to provide direct answers to the user's queries.}
	\label{fig:ei_overview}
\end{figure}

\section{Introduction}

Recent advancements in large language models (LLMs)~\cite{ouyang2022training, achiam2023gpt,dubey2024llama} have transformed user interactions with knowledge-based systems by enabling chatbots to synthesize vast amounts of information, surpassing human cognitive limits. 
While traditional search engines like Bing~\cite{bingsearch} and Google~\cite{googlesearch} efficiently address straightforward queries like fact-checking or retrieving specific information, LLM-based chatbots truly shine in guiding users through more open-ended and exploratory tasks~\cite{Graphologue, Tang24pdf, Wu22prompt, Kim2024datadive, chi24google}. For instance, they can support users in researching emerging scientific fields~\cite{Steven24reserch}, or planning intricate projects that require synthesizing information~\cite{Sensecape} from diverse sources and close user engagement. This capability has made LLM-based chatbots invaluable in helping users navigate unfamiliar areas.

Despite their extensive capabilities, LLM-based chatbots face challenges in delivering personalized assistance during exploratory tasks~\cite{iui24chatgptdsats, Jean2023chatgpt}, particularly when they lack access to user-specific data, such as past interactions and personal preferences. This limitation forces them to rely heavily on user-provided inputs for personalization. In exploratory tasks, where users often begin with vague queries, generating answers based solely on limited information typically results in generic and impractical recommendations. For example, in travel planning, a query like ``\textit{Plan a 5-day trip to Hawaii}'' without additional details (e.g., budget, preferred activities, or accommodation preferences) leads to an overly generic response, as shown in the bottom part of Figure~\ref{fig:ei_overview}. This lack of personalization arises either because users may not be aware of the essential details to include or because providing detailed information imposes cognitive load on users. Consequently, the travel plan generated from such a vague query would be insufficiently tailored to the user's specific needs. While users can refine their requests after receiving an immediate solution from the initial vague query, limitations remain, such as bias toward existing solutions and reduced motivation to explore alternative dimensions, and dissatisfaction is not always resolved despite using specification tactics~\cite{iui24chatgptdsats}.

In this study, we introduce the \textbf{C}ollaborative \textbf{A}ssistant for Pe\textbf{r}sonalized \textbf{E}xploration (\textbf{CARE}) system to tackle the challenges associated with generating personalized solutions in exploratory tasks. CARE facilitates tailored exploration through a user interface composed of three key elements (see Figure~\ref{fig:ei_overview}): the \textit{Chat Panel}, \textit{Solution Panel}, and \textit{Needs Panel}, enabling dynamic user interaction and iterative solution refinement. The system operates on a multi-agent collaboration framework featuring serveral LLM-driven agents responsible for managing the entire process, from user needs identification to solution generation. These agents collaborate to extract both explicit and implicit user needs, prioritize inquiries, and craft customized solutions. The system also ensures transparency, flexibility, and usability, allowing users to refine their queries iteratively and receive personalized, actionable results.

To evaluate the effectiveness of the proposed CARE system, we performed a within-subject user study with 22 participants. Each participant completed two exploratory tasks using both the CARE system and a baseline LLM-based chatbot. The results revealed a clear preference for CARE, with 16 out of 22 participants favoring it over the baseline. Participants praised CARE for reducing cognitive load, inspiring creative exploration, and providing more personalized solutions. Many appreciated how CARE’s structured interface helped them organize complex tasks systematically and encouraged them to think about aspects they had not considered. Overall, CARE was seen as more engaging and effective in supporting exploratory tasks compared to the baseline system.

Our contribution can be summarized as follows:
\begin{itemize} 
\item We introduce a collaborative, chat-based interface designed to facilitate users in accomplishing exploratory tasks, particularly when they initially present vague or ambiguous queries due to insufficient knowledge.
\item The interface is powered by an LLM-based multi-agent collaboration system, demonstrating its functionality and effectiveness in aiding users.
\item We conduct a comprehensive user study to evaluate the effectiveness of the proposed system, providing valuable insights that can inform the design of future user-centric exploratory systems. 
\end{itemize}

\section{Related Work}

\subsection{Large Language Models for Exploratory Tasks}
LLM-based human-computer interaction (HCI) systems, such as ChatGPT~\cite{ChatGPT}, Copilot~\cite{Copilot}, and Gemini~\cite{Gemini}, have expanded the role of AI in supporting users with exploratory tasks. Unlike traditional search engines that excel at handling well-defined queries, LLMs demonstrate notable strengths in assisting with open-ended, complex tasks. Recent work in HCI has explored the integration of LLMs into various domains, including creative workflows~\cite{Xu2024Jamplate, Wang24LAVE, Tang24pdf, Wu22prompt}, help-seeking~\cite{iui24software, Qian24prompt}, planning~\cite{zhang2024ask, chi24google}, and data exploration~\cite{Kim2024datadive}, revealing the potential of LLMs to facilitate adaptive interactions in complex contexts. Much of this research focuses on advancing prompt-based methods to enhance LLM performance, unlocking reasoning, problem-solving, and planning capabilities~\cite{Brown20prompt, Jules23prompt}. While effective, prompt-based techniques encounter challenges such as inconsistent instruction-following~\cite{Zamfirescu23herd, Zamfirescu2023prompt, Jang2024chatgpt}, leading to reliability issues in chatbot behavior. These limitations underscore the need for systems that can more effectively manage the complexity and ambiguity inherent in exploratory tasks.

Our work builds on this body of research by introducing a multi-agent collaboration system specifically designed for exploratory tasks. Task decomposition, a method effective in managing complexity, plays a central role in our approach. While some systems attempt to address exploratory tasks by prompting a single LLM agent, we found that a multi-agent architecture offers more rigorous task management and better alignment with user needs. Single-agent system often struggle to maintain focus across multiple responsibilities or lose track of ongoing tasks~\cite{iui24chatgptdsats, Wu22prompt}, which can disrupt user progress. By assigning distinct roles to each agent, our system maintains greater control over task decomposition, milestone tracking, and user intent discovery, resulting in a more structured and personalized exploratory experience.


\subsection{Personalization and Inspiration in LLM-Based Systems}
Personalization is a critical factor in enhancing user experience across AI-driven systems~\cite{Laban20personalize, Vossen24personalize}. Many interactive platforms, such as recommendation engines~\cite{Silva24recommender, Baek24recommender} and adaptive learning systems~\cite{Nazari24personalize, Li24adaptive}, achieve personalization by leveraging user data, including past behavior, preferences, and explicit feedback. However, in the context of LLM-based systems, personalization presents greater challenges, as these models typically lack access to user-specific data unless explicitly provided within the conversation (\eg, via prompts)~\cite{Jean2023chatgpt}. Consequently, LLMs often generate responses based only on the limited context of the current interaction, which can result in generic or irrelevant outputs, particularly in exploratory tasks. This passive approach to personalization—where LLMs react solely to the user's inputs and user follows up with new prompt—often leads to incomplete articulations of the user's needs, resulting in less comprehensive and relevant responses~\cite{Hyunwoo24passive}. This is especially problematic in open-ended, complex tasks where users may not fully express or even realize their own requirements.

It is important to note that while adding a memory module to LLM-based chatbots can improve personalization by enabling the system to reference prior interactions, it may take a significant number of exchanges to gather enough data to meaningfully enhance this personalization. In the early stages of interaction, the chatbot may lack relevant insights, limiting its ability to provide tailored responses. Additionally, for new or unfamiliar tasks, previously stored information may not be useful, as exploratory tasks often require fresh context that past exchanges cannot fully provide.

Inspiration is another critical component, particularly in open-ended tasks. Prior research has shown that LLMs can stimulate creativity, assisting users in generating new ideas or perspectives~\cite{Gero22inspire, Petridis24inspire}. Unlike studies focused on sparking creativity or employing Socratic questioning in educational contexts~\cite{Chang23Socratic, Ding24Socratic}, our work emphasizes the discovery of users' implicit needs and requirements. Rather than merely responding to explicit queries, we employ proactive inquiry techniques that encourage users to think expansively. By actively probing for underlying goals and unarticulated needs, our system helps users consider aspects they may not have initially thought of, ultimately leading to more comprehensive and personalized outcomes.

\subsection{User Interface Support for Thinking and Task Organization}
Traditional chatbots and most LLM-based conversational systems rely on linear, dialogue-based interfaces, generating responses in a text stream without exploiting graphical user interfaces (GUIs). However, recent work has begun to explore the potential of graphical interfaces to better support users in complex tasks. For example, Sensecape integrates LLMs into an interactive system to facilitate information foraging and sensemaking~\cite{Sensecape}, while Graphologue converts text responses into interactive diagrams to enhance information-seeking and question-answering tasks~\cite{Graphologue}. ExploreLLM employs a card-based schema to help users structure their thoughts and explore alternative solutions in a more organized manner~\cite{chi24google}. A key challenge with linear text interfaces is the difficulty in locating and recalling relevant information, as important details can become buried within lengthy dialogue streams~\cite{Popolov00Conversation, iui24chatgptdsats, Goodwin2015linear}. Our work contributes to this area by introducing a novel conversational interface, which separates the solution display from the dialogue interactions, while incorporating a dedicated section for managing user requirements. This structured design reduces cognitive load, enabling users to focus more easily on their tasks and better manage information flow during extended interactions.

\section{Motivating Insights from Empirical Studies on Chat-Based Interfaces}
Our proposed system is motivated by challenges users face when interacting with conventional LLM-based chatbots. 
A review of existing research on systems like ChatGPT~\cite{iui24chatgptdsats, Borji2023chatgpt, Floridi2023chatgpt, Sukhpal2023chatgpt, Partha2023chatgpt, Jang2024chatgpt, Morteza2023chatgpt}, highlights key issues that persist in chat-based systems, especially regarding personalization~\cite{Partha2023chatgpt}, relevance~\cite{Jang2024chatgpt}, and sustained contextual understanding~\cite{Morteza2023chatgpt} in complex, open-ended scenarios.

Several studies~\cite{iui24chatgptdsats, Jean2023chatgpt, qian2024tell, Deng2023intent} have consistently identified user dissatisfaction stemming from LLMs’ frequent inability to capture the nuanced intent behind vague or incomplete queries. Users often encounter responses that are generic, irrelevant, or redundant, which diminishes trust in the system, particularly in tasks requiring deeper exploration or personalization. For example, research shows that users seeking localized information about a service may struggle when systems fail to interpret location-specific terms or understand personal preferences~\cite{zhang2024ask}, leading to frustration and ineffective recommendations. This problem is further exacerbated in exploratory tasks, where users themselves may not have well-formed goals and rely on the system to guide them through ambiguity.

Moreover, existing work~\cite{iui24chatgptdsats, Zamfirescu2023prompt, Jiang22prompt, Wu22prompt} has revealed that many users feel ill-equipped to manage the iterative, multi-turn interactions necessary to refine system outputs in such scenarios. The cognitive load imposed by these systems can be high, particularly when users are required to repeatedly clarify or reframe their inputs. Even with prompt engineering and optimization techniques designed to enhance LLM performance, users still frequently experience dissatisfaction when systems cannot offer personalized assistance based on minimal input or evolving needs.

From these insights, several design goals emerge as essential for creating more collaborative chat-based interfaces:

\begin{itemize}
    \item \textbf{Improved Contextual Understanding}: System must be capable of maintaining and interpreting user intent over multiple turns, adapting to ambiguous or evolving user inputs to provide more relevant guidance.
    \item \textbf{Enhanced Personalization}: System should be designed to tailor responses dynamically, accommodating user preferences and contextual needs, even when the initial input lacks specificity.
    \item \textbf{Facilitation of Inspired Exploration}: In addition to providing personalized solutions, the system should inspire users to think implicit needs or aspects, promoting comprehensiveness in open-ended tasks.
    \item \textbf{Reduction of Cognitive Load}: To mitigate the frustration associated with managing complex tasks, the interface should provide structure by organizing information effectively and helping users refine their queries.
\end{itemize}
These design goals are informed by extensive literature on user interaction with LLMs, providing a clear direction for improving the CARE system’s capabilities in supporting personalized, exploratory tasks.

\section{CARE: Collaborative Assistant for Personalized Exploration} 

\begin{figure}[!htb]
	\centering
	\includegraphics[width=\textwidth]{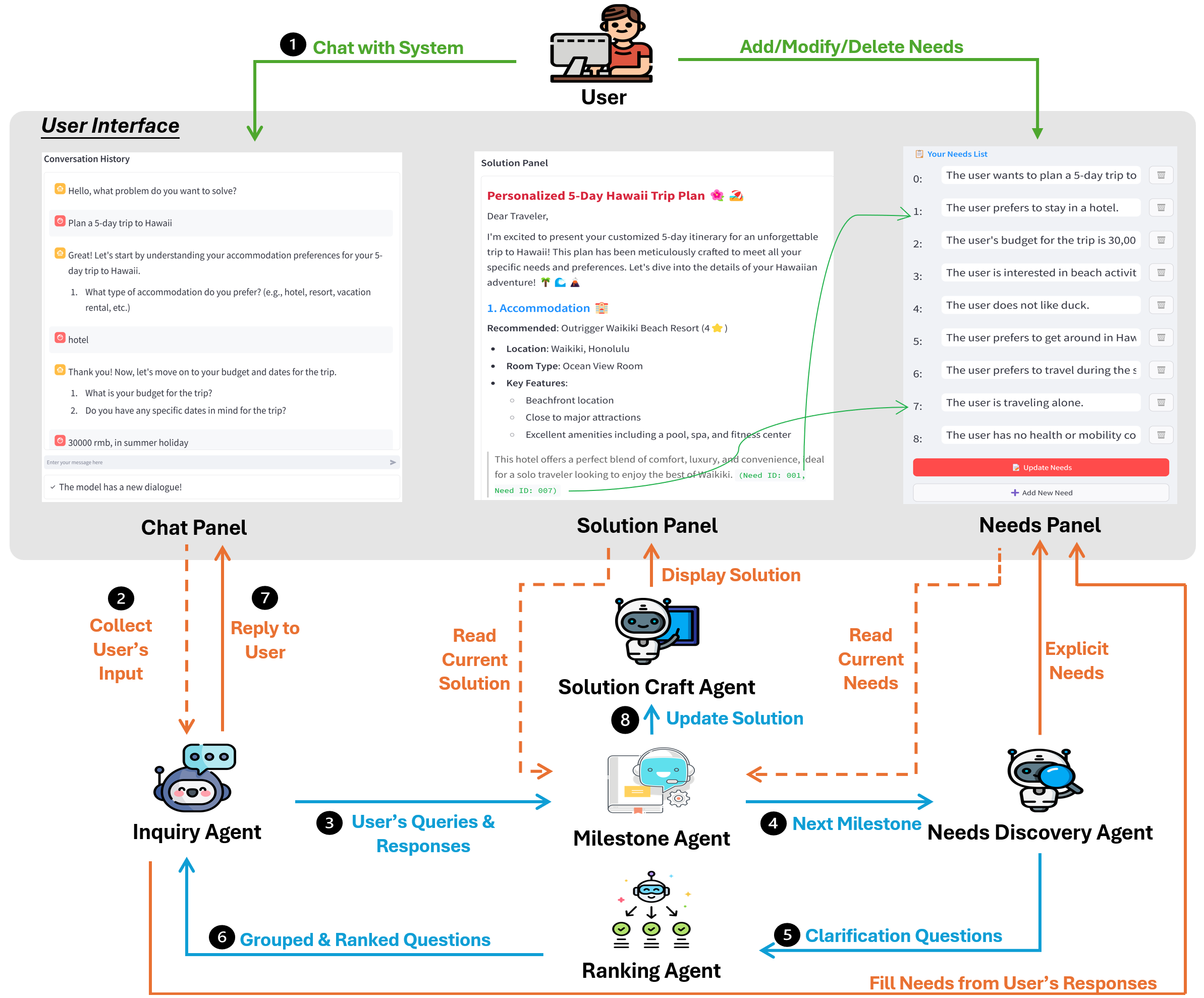}
	\caption[Figure]{Overview of the CARE system. The gray area represents the User Interface, where users interact through the Chat, Solution, and Needs Panels. At the bottom, CARE's back-end consists of several agents, including the Inquiry Agent, Needs Discovery Agent, Solution Craft Agent, Milestone Agent, and Ranking Agent, which collaborate to process user inputs and generate personalized solutions. \textcolor[HTML]{4da72e}{$\rightarrow$} represents user interactions, such as chatting or updating needs. \textcolor[HTML]{119ed5}{$\rightarrow$} represents the internal data flow between agents. \textcolor[HTML]{e97033}{$\rightarrow$} represents that the agents write data to the interface. \textcolor[HTML]{e97033}{$\dashrightarrow$} represents that the agents retrieve data from the interface. 
 }
	\label{fig:ei_sys_fig}
\end{figure}

In this section, we describe the proposed system, \textbf{C}ollaborative \textbf{A}ssistant for Pe\textbf{r}sonalized \textbf{E}xploration (\textbf{CARE}), which aims to achieve the design goals derived in the previous section. The overall system architecture of CARE is illustrated in Figure~\ref{fig:ei_sys_fig}, and the subsequent sections elaborate on the user interface and back-end components.

\subsection{User Interface}

As illustrated in Figure~\ref{fig:ei_sys_fig}, the CARE's user interface consists of three main panels: the \textit{Chat Panel}, \textit{Solution Panel}, and \textit{Needs Panel}. 
Next, we illustrate their functions using a travel planning scenario.

The \textit{Chat Panel} functions as the primary interface where users interact with CARE, entering initial queries such as ``\textit{Plan a 5-day trip to Hawaii}''. Through iterative feedback and conversation with the chatbot, users refine their requests, resulting in a dynamic multi-turn dialogue.

As the conversation progresses, CARE generates or updates a solution tailored to the user’s needs, which is displayed in the \textit{Solution Panel}. To enhance user experience, the solution is presented in Markdown format with rich text features (e.g., tables, emojis) for improved readability. In the travel planning case, the solution might include recommendations for accommodation, transportation, and other essential details, ensuring clarity and engagement.

The \textit{Needs Panel} complements this by explicitly displaying the collected user needs and identifying any points requiring clarification. This allows the system to ``ground'' the solution shown in the \textit{Solution Panel} by linking specific parts of the solution to particular user needs.
For example, as shown in Figure~\ref{fig:ei_sys_fig}, the green text of ``\textit{Need ID: 001}'' in the \textit{Solution Panel} highlights how parts of the travel plan address needs listed in the \textit{Needs Panel}, ensuring traceability and transparency between user preferences and the generated plan. Additionally, users can modify, add, or delete needs directly in the \textit{Needs Panel} for greater flexibility.

In summary, the CARE interface introduces a novel design that separates the solution display in the \textit{Solution Panel} from the conversational interactions in the \textit{Chat Panel}, while incorporating a dedicated \textit{Needs Panel} to manage and cross-reference user needs within the solution.

\subsection{LLM-Powered Multi-Agent Collaboration System}

The back-end of the CARE system is powered by an LLM-driven multi-agent collaboration framework, as depicted in the bottom part of Figure~\ref{fig:ei_sys_fig}. It comprises five specialized LLM-based agents: the \textit{Inquiry Agent}, \textit{Milestone Agent}, \textit{Needs Discovery Agent}, \textit{Ranking Agent}, and \textit{Solution Craft Agent}. These agents collaborate with the user to progressively manage the solution development across three panels in the UI, discovering implicit needs and generating personalized solutions. While similar systems can be created by prompting a single LLM-based agent, we found that a multi-agent approach ensures more rigorous task management and better alignment with our design goals. Single-agent models can lose track of tasks or become overwhelmed by managing too many responsibilities. By assigning distinct roles to each agent, we maintain better control over individual components, ensuring consistency in task decomposition, milestone tracking, and user need discovery. Next, we introduce the overall interaction workflow of this multi-agent system, using the same example case of \textit{``Plan a 5-day trip to Hawaii.''}

\subsubsection{Interaction Workflow among Agents}
The workflow, shown in Figure~\ref{fig:ei_sys_fig}, is triggered by the query initiated by the user \circlednum{1} in the \textit{Chat Panel}. Subsequently, the \textit{Inquiry Agent} receives the user's request \circlednum{2} and passes it to the \textit{Milestone Agent} \circlednum{3}, who assesses if enough user needs are collected to proceed. If more details are needed, a new milestone is set \circlednum{4}, such as \textit{``understand user’s hotel preferences.''} The \textit{Needs Discovery Agent} then extracts  needs from the user’s input and generates follow-up questions to uncover additional needs that not mentioned by user \circlednum{5}.

The \textit{Ranking Agent} organizes these questions, grouping and prioritizing them \circlednum{6} before they are presented to the user. These questions are then passed back to the \textit{Inquiry Agent}, who presents them to the user in an intuitive manner \circlednum{7}. As the user responds, their answers are populated into the \textit{Needs Panel}. This process continues iteratively until the \textit{Milestone Agent} confirms that the collected needs are sufficient, prompting the \textit{Solution Craft Agent} to generate a personalized solution \circlednum{8}, which is then presented in the \textit{Solution Panel}.

Upon receiving the solution, the user can choose to refine the solution in two ways. They can directly communicate new preferences to the \textit{Inquiry Agent}, which may trigger additional needs discovery steps, or if the preference is straightforward, the \textit{Milestone Agent} will pass the new feedback directly to the \textit{Solution Craft Agent} for immediate integration into a revised solution. Alternatively, the user can manually modify, add, or delete items in the \textit{Needs Panel}. Any such updates will prompt the \textit{Solution Craft Agent} to generate a new solutions based on the latest needs.

The above workflow illustrates how the agents collaborate to deliver solutions tailored to users' preferences while facilitating a comprehensive needs discovery process. The following sections describe each agent in detail.

\subsubsection{Inquiry Agent}
The \textit{Inquiry Agent} is designed to facilitate smooth user interactions by gathering and clarifying user needs in an intuitive manner. One of its key responsibilities is refining the questions generated by the \textit{Needs Discovery Agent} before they are presented to the user. This process involves simplifying the language of questions and providing default options, which significantly reduces the user's cognitive load. For instance, instead of presenting a broad question like, \textit{``What kind of accommodation do you prefer?''}, it offers more focused choices, such as \textit{``Hotel or Airbnb?''}, thereby streamlining the decision-making process for users.

In addition to refining questions, the \textit{Inquiry Agent} actively assists users who may struggle with specific questions or express uncertainty. It provides pertinent explanations or definitions, such as explaining ``Airbnb'', enabling users to make well-informed decisions. Moreover, once the user responds, the \textit{Inquiry Agent} meticulously records the answers to the \textit{Needs Panel}, ensuring that all user preferences are accurately captured for future stages of the process.

\subsubsection{Milestone Agent}
The \textit{Milestone Agent} plays the strategic coordinator role in CARE by determining the next critical steps (i.e., milestones) based on the existing user’s input and collected needs, ensuring that all necessary user needs are gathered before moving forward with solution generation. 

The \textit{Milestone Agent} has two primary responsibilities. First, if it determines that more specific user needs are required or the user requests an improvement on the existing solution, it sets the next milestone. This milestone typically involves collecting missing or incomplete information necessary for building a comprehensive solution. The \textit{Milestone Agent} uses the current \textit{Needs Panel} and input from the user to establish these milestones, ensuring that they are specific, actionable, and contribute directly to solving the user’s problem. It references previously established milestones to avoid redundancy and ensures each new milestone is unique and focused on specific, measurable outcomes. Additionally, the agent carefully considers dependencies between tasks, breaking down complex problems into manageable milestones that the CARE system can address step by step.

Second, if the recorded user needs are sufficient, the \textit{Milestone Agent} notifies the \textit{Solution Craft Agent} to begin generating a solution. At this stage, no new milestones are created, and the process transitions to solution generation. This decision is based on whether the current \textit{Needs Panel} fully addresses the user’s query.

In cases where the user manually updates their needs via the user interface, the \textit{Milestone Agent} immediately passes this information to the \textit{Solution Craft Agent}, triggering a plan update without setting a new milestone. This ensures flexibility in how the CARE system adapts to user input and allows for rapid response to direct feedback.

\subsubsection{Needs Discovery Agent}
The \textit{Needs Discovery Agent} serves a vital function in identifying and documenting user needs, both explicit and implicit, throughout the interaction with the system. This process involves two critical tasks that ensure comprehensive coverage of user Needs.

First, the agent is responsible for extracting explicit needs directly from the user's input. Explicit needs are those that the user clearly articulates. For instance, if a user expresses a desire to \textit{``Plan a 5-day trip to Hawaii,''} the agent will identify and extract specific needs, such as \textit{``the destination is Hawaii''} and \textit{``the trip duration is 5 days''}. These explicit needs are systematically recorded in the \textit{Needs Panel} to ensure they are fully documented, making it simple for users to review their own needs, while also allowing the CARE system to further develop personalized solutions.

Second, the agent plays an essential role in identifying implicit needs, which are often unspoken but critical to achieving the current milestone. Implicit needs are inferred based on the chat history and the specific demands of the milestone. For example, when the task involves selecting a hotel, the agent may infer preferences such as whether the hotel needs to be close to popular attractions or if the user has specific needs for certain amenities, such as a gym, swimming pool, or free Wi-Fi. By anticipating these needs, the agent ensures a more complete understanding of the user's needs. Once the agent has completed its inference, it generates clarification questions to further refine the user's preferences. These questions may include, for example, \textit{``whether the user has specific requirements for the accommodation's location''} or \textit{``any preferences regarding hotel amenities.''} The generated questions are then added to the \textit{Needs Panel}, although invisible on the UI to avoid user distraction.

\subsubsection{Ranking Agent}
The \textit{Ranking Agent} brings structure and order to the system by organizing and prioritizing the clarification questions identified during the user needs discovery process. Instead of overwhelming users with a disjointed list of queries, the agent groups related questions thematically, such as accommodation preferences or budget constraints, allowing the user to focus on one topic at a time. This approach not only minimizes cognitive overload but also keeps the interaction flowing smoothly.

Once the questions are grouped, the \textit{Ranking Agent} further arranges them in a logical sequence. It starts with simpler, more general inquiries, then gradually progresses to more detailed, specific questions. By controlling the pacing and order of the interaction, the agent ensures that users can answer with confidence and clarity, without being rushed or overwhelmed. This results in a seamless experience where the user is led through the interaction with ease, while all necessary information is gathered efficiently.

\subsubsection{Solution Craft Agent}
The \textit{Solution Craft Agent} is central to transforming user needs into concrete, personalized solutions. This agent excels at turning complex sets of needs into clear, actionable outcomes. After analyzing the user’s preferences and constraints, it organizes the solution in a way that is both transparent and user-friendly. Each solution is neatly structured, with headings and concise explanations, making it easy for users to follow and understand how their specific needs are being addressed. By linking every recommendation to a particular need, the agent ensures that users can see exactly how their input shapes the proposed solution.

Personalization is another key strength of the \textit{Solution Craft Agent}. Rather than offering generic solutions, it tailors each recommendation to align with the user’s context, preferences, and limitations. Whether suggesting accommodations that fit within a specific budget or transportation options that align with the user’s schedule, the agent ensures that the solution is relevant and practical. This emphasis on customization guarantees that each user receives a solution that is uniquely suited to their situation, making the final outcome both useful and highly personalized.

\subsection{Implementation Details}

The CARE system is a web-based application featuring a UI developed with Streamlit~\cite{Streamlit} and a back-end multi-agent system constructed using AutoGen~\cite{wu2023autogen}. Additionally, we implemented a conventional LLM-based chat interface by directly prompting the base LLM, which serves as the baseline system for our user study. Both CARE and the baseline systems utilize GPT-4o as the underlying LLM, with all corresponding prompts detailed in Supplementary Material.


\section{User Study}
\label{sec:user_study}

To minimize the effects of participant variability, we conducted a within-subject user study with 22 participants, comparing CARE to a baseline LLM-based chatbot as described above. Each participant completed two exploratory tasks, travel planning and skill learning, using both systems. To control for order effects, we counterbalanced the order of the systems and tasks, resulting in four settings. The study aimed to address the following research questions (RQs):
\begin{itemize}
\item \textbf{RQ1.} How does the CARE system compare to the baseline system in supporting users in exploratory tasks in terms of interaction experience and user satisfaction?
\item \textbf{RQ2.} How does the CARE system provide personalized and inspired solutions compared to the baseline system? 
\item \textbf{RQ3.} How does the CARE system’s thoughtful interface design help reduce cognitive load during complex, open-ended tasks, compared to the baseline system?
\end{itemize}




\subsection{Procedure}
Each study session began with an introduction to both systems, during which participants received basic instructions on how to interact with the systems (e.g., explanation on UI layout, reminder on no web search function). Participants then completed a demographic questionnaire (Section~\ref{sec:appendix_survey1}) detailing their background and prior experience with LLM-based chatbots. Each participant completed two exploratory tasks using both systems. We encouraged participants to spend 5-10 minutes on each task. After interacting with each system, participants filled out a post-task questionnaire (Table~\ref{tbl:questionnaire}, questionnaire format in Section~\ref{sec:appendix_survey2}) to evaluate their experience. Throughout the session, participants were encouraged to think aloud. Follow-up interviews (Section~\ref{sec:appendix_interview}) were conducted at the end of each session to gather additional qualitative insights, particularly focusing on any notable highlights or frustrations with each system.

Each session lasted approximately one hour. 
The study was conducted in person, and participants were compensated with cash equivalent to two hours of the local minimum wage in appreciation of their time.



\subsection{Data Collection and Analysis}

To evaluate how effectively CARE supports users in exploratory tasks, we designed a comprehensive questionnaire (Table~\ref{tbl:questionnaire}) focusing on several key aspects of the user experience: Interaction, Cognitive Load, Inspiration, Comprehensiveness, and Personalization. During each study session, we recorded participants’ screen activities and audio to capture their think-aloud verbalizations, allowing us to gather qualitative insights alongside quantitative data.

For the quantitative analysis, we employed a combination of statistical methods, including paired t-tests and Pearson’s Chi-Square tests, to assess the significance of our findings. The paired t-test allowed us to determine whether there was a statistically significant difference in user ratings between CARE and the baseline system. The Chi-Square test was used to examine categorical differences in participant responses across various dimensions of the user experience. \xt{@Yingzhe help to review}
To provide further context for the quantitative results, we analyzed the think-aloud data and post-task interviews using an inductive qualitative approach. This allowed us to probe into the reasons behind the participants’ ratings and explore their subjective experiences with the two systems. The qualitative feedback was instrumental in corroborating the experimental findings and providing deeper insight into how CARE facilitated user exploration, reduced cognitive load, and supported creativity.

\begin{table}[!ht]
    \centering
\begin{tabular}{l|l}
\toprule
\multicolumn{1}{c|}{\textbf{Measure}} & \multicolumn{1}{c}{\textbf{Statement (5-Point Likert Scale)}} \\ \hline
\textbf{Interaction} & Q1. I enjoy the way I interact with the system. \\
\textbf{Cognitive Load} & Q2. The system’s UI helps me organize complex tasks and reduces my cognitive load.\\ 
\textbf{Inspiration} & Q3. Interacting with the system inspires me to consider new aspects of exploratory tasks.\\
\textbf{Comprehensiveness} & Q4. The answers provided by the system feel comprehensive and sufficient to me.\\
\textbf{Personalization} & Q5. The answers provided by the system match my personal needs.\\
\bottomrule
\end{tabular}
\caption{Post-task questionnaire filled out by participants after they interacted with two systems, one with CARE and the other with the Baseline. Each statement was rated on a 5-point Likert scale (the larger the better).}
\label{tbl:questionnaire}
\end{table}

\subsection{Participants}
We recruited 22 participants (4 female, 18 male) through a call for participation distributed via the university's mailing list, as well as through word-of-mouth referrals. Participants represented a diverse range of academic disciplines, including Computer Science (9), Engineering (8), Natural Sciences (2), Social Sciences (2), and Arts (1). Their educational backgrounds varied, with 4 undergraduates, 17 master’s students, and 1 PhD student.

All participants had prior experience with conversational LLM-based systems, such as ChatGPT, but their frequency of use varied. Seven participants reported using such systems less than once per week, five used them several times per week but not daily, and ten were frequent users who interacted with LLM-based chatbots multiple times per day.

\section{Quantitative Results}
\begin{figure}[!t]
	\centering
	\includegraphics[width=1\textwidth]{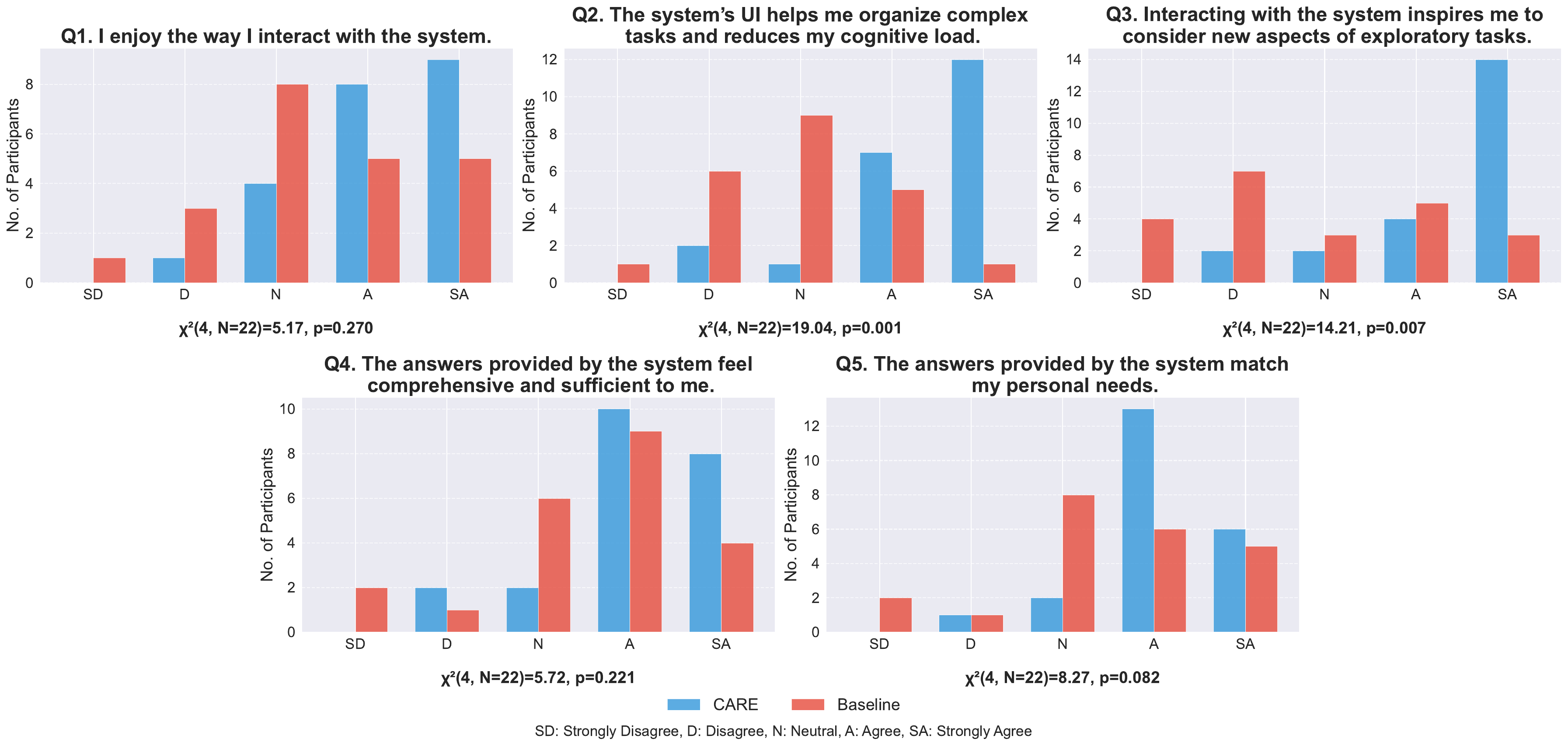}
	\caption[Figure]{Comparative analysis of user responses to the CARE and baseline systems across five key aspects of user experience.}
	\label{fig:questionnaire}
\end{figure}
The binary choice results, collected through post-task interviews, reveal a clear preference for CARE over the baseline system, with 16 out of 22 participants favoring CARE. This finding aligns with the quantitative data from the five survey questions listed in Table~\ref{tbl:questionnaire}, as depicted in Figure~\ref{fig:questionnaire}. When averaging scores across all five questions, CARE was rated significantly higher ($t = 3.07, p = 0.0058$), with an average score of 4.20 ($\sigma = 0.78$) compared to the baseline's 3.25 ($\sigma = 0.91$). Cohen’s $d = 1.12$ also illustrates the considerable difference in perceived effectiveness, highlighting the participants' clear preference for CARE. We now turn to the detailed statistical analysis of each survey question.



\noindent\textbf{Overall Interaction Experience.} The Pearson Chi-Square analysis ($\chi^2(4) = 5.17, p = 0.27$) indicates no significant difference in overall satisfaction between the two systems in Q1. However, a larger proportion of participants rated CARE favorably in the ``Agree'' and ``Strongly Agree'' categories, with 17 out of 22 participants finding CARE engaging, compared to only 10 for the baseline. This pattern, though not statistically significant, suggests that CARE might provide a better interaction experience for the majority of users.

\noindent\textbf{Reducing Cognitive Load.} According to the results in Q2, users perceived that CARE significantly reduced cognitive load in organizing and managing complex tasks compared to the baseline ($\chi^2(4) = 19.04, p = 0.001$). Notably, 13 participants strongly agreed that CARE was more systematic in presenting and organizing information, compared to only one participant in the baseline system.

\noindent\textbf{Inspiring New Ideas.} The Chi-Square result for Q3 (\( \chi^2(4) = 14.21, p = 0.007 \)) demonstrates a statistically significant difference between the two systems. Notably, 14 participants strongly agreed that CARE inspired them to generate new ideas or explore alternative perspectives during the interaction, compared to only 5 participants in the baseline system. This demonstrates that CARE encourages a more exploratory and creative interaction compared to the baseline.

\noindent\textbf{Solution Comprehensiveness.} The Chi-Square analysis of Q4 (\( \chi^2(4, N=22) = 5.72, p = 0.221 \)) reveals no significant difference between the two systems in terms of solution comprehensiveness. This outcome is encouraging, as it implies that CARE offers tailored solutions without compromising the comprehensiveness of the responses.


\noindent\textbf{Solution Personalization.} CARE showed a notable advantage (\( \chi^2(4) = 8.27, p = 0.082 \)) in terms of solution personalization (Q5), with 19 participants agreeing that CARE's plans were more personalized, in contrast to 11 for the baseline system. While the $p$ value did not meet the conventional threshold of $0.05$, it suggests a trend toward significance, indicating potential for future enhancements to the CARE system.

The above quantitative analysis highlights a clear preference for CARE over the baseline system, with most participants favoring CARE in terms of effectiveness and user experience. While not all survey questions showed statistically significant differences, CARE consistently outperformed the baseline in reducing cognitive load, inspiring new ideas, and providing personalized solutions. These findings set the stage for a deeper exploration of qualitative insights from the post-task interviews, where participants' nuanced perspectives on their interaction with CARE will be discussed.

\section{Findings}

The findings from our user study are organized around addressing the three research questions listed in Section~\ref{sec:user_study}, providing insights into the strengths of the CARE system compared to the baseline. First, we explore how CARE’s inquiry process supports exploratory tasks by fostering deeper exploration, ensuring comprehensiveness, and delivering personalized outputs, aligning with \textbf{RQ1} and \textbf{RQ2}. Next, we examine how CARE’s collaborative workspace effectively organizes user needs, generated solutions, and chat interactions, contributing to reduced cognitive load and improved task performance, directly addressing \textbf{RQ3}. These findings collectively highlight the key benefits of CARE’s design in enhancing user experience and productivity.

\subsection{CARE’s Inquiry Process Inspires Exploration, Ensures Comprehensiveness, and Personalizes Outputs}

\subsubsection{CARE’s Inquiry Process Inspires Users by Uncovering Implicit Needs} 
Through our investigation, we found that CARE effectively uncovers users' implicit needs by encouraging them to think broadly and respond to proactive inquiries. During the interviews, we further identified that users' implicit needs can be divided into two distinct categories: \textit{hidden needs} and \textit{latent needs}. \textit{Hidden needs} refer to those needs that users are aware of but do not explicitly express, while \textit{latent needs} are those that users do not recognize until they are prompted by CARE’s exploratory questions. 

CARE effectively inspires users to consider both hidden and latent needs. Specifically, most participants (17/22) indicated that CARE’s questioning strategy encouraged them to think about aspects they would not have otherwise considered. For instance, \textit{P7} highlighted how CARE helped him uncover hidden needs while planning a trip: \textit{``For example, when planning a trip, it (CARE) would ask me if I prioritize sightseeing, food, or shopping, and if I have a preference, it can better tailor the itinerary for me.''} This kind of questioning reveals needs that the user might be aware of but did not communicate explicitly. On the other hand, \textit{P19} emphasized how CARE brought to light latent needs during the process of creating a baseball training plan: \textit{``I think System A (CARE) inspired me a lot. For example, when I wanted to start playing baseball, it asked about the coach’s style and how much time I wanted to train—especially questions about the coach’s style, which I had never thought about before.''} These examples demonstrate how CARE’s inquiry process guides users to explore crucial factors they initially overlooked but which are critical to making more informed decisions. 

In contrast, participants found that the baseline system reacted passively, only responding to explicit inputs without offering additional considerations. \textit{P11} pointed out: \textit{``B (Baseline) only answered what I asked, but A (CARE) asked about things like my budget and dietary preferences, which I hadn’t thought about.''} This passive approach often led to incomplete articulation of the user’s needs, resulting in less comprehensive and relevant output. 


\subsubsection{CARE's Milestone-Driven Approach Improves User Focus And Personalizes Task Progression.} 
To ensure that CARE continuously focuses on the user's current needs, tasks are organized into sequential milestones by \textit{Milestone Agent}, guiding other agents in the system step by step through the planning process. Each milestone represents a distinct phase of progress, enabling agents to focus on one task at a time while maintaining a coherent long-term strategy. As users provide feedback or refine their goals based on the current milestone, the system dynamically updates the plan or generates the next milestone, ensuring continuous alignment with the user's evolving needs. 

This milestone-driven approach was noticed and appreciated by participants (10/22). \textit{P20} described this as, \textit{``It (CARE) asked questions that helped me break down a big problem into smaller, more systematic tasks. This way, I could understand my needs better, and the system could offer more personalized settings''}. Similarly, \textit{P8} mentioned, \textit{``The system A (CARE) would update after each step, and I could see how my decisions affected the plan, which helped me stay on track.''}



The baseline system, however, lacked this granular task breakdown.
Without the ability to iteratively refine and adjust based on continuous feedback, users found it difficult to maintain focus on evolving goals. 
\textit{P15} remarked, \textit{``With the another system (Baseline), I often felt lost trying to manage everything at once.''} 


\subsubsection{Comprehensive Inquiry Process Ensures More Detailed, Structured, and Complete Solution.}
CARE ensures that users' needs are comprehensively covered through a systematic inquiry process during interactions. This comprehensiveness is reflected in two dimensions: \textit{the comprehensiveness of the inquiry process} and \textit{the comprehensiveness of the generated solution}. The former refers to the system’s ability to cover multiple aspects of the user's needs through its multi-dimensional questions, while the latter pertains to the final solution being more structured and complete. 

Some participants (13/22) expressed a preference for this proactive, multi-dimensional inquiry-driven interaction style. For example, \textit{P12} noted: \textit{``A (CARE) asks follow-up questions based on your answers and helps fill in things you might not have thought about. I think it’s very comprehensive.''} Similarly, \textit{P4} highlighted CARE’s ability to expand a single question into multiple dimensions, noting: \textit{``A (CARE) is suitable for general users because many times you only have one question in your mind, but it can help you break it down into various aspects that need to be considered.''} 

Moreover, CARE incorporates a \textit{Ranking Agent}, which prioritizes questions to allow users to first answer simpler, foundational questions, thereby reducing cognitive load during interactions. This design ensures that questions are presented in a structured sequence, avoiding users being overwhelmed by excessive information at once. 
As P8 stated: \textit{``It’s (CARE's) interactive, and it asks simple questions, like multiple-choice or short questions. I felt relaxed while interacting with it, and it was really good at recognizing my intent and adding that to the list of needs.''}



This level of comprehensiveness during interaction was lacking in the baseline system.
Users had to think through the task and provide as much information as possible upfront by themselves, and the system did not proactively ask about additional dimensions of the task. 
Some participants (8/22) expressed frustration with this limitation. \textit{P21} explained: \textit{``With chatbot B (Baseline), I had to ask more follow-up questions myself to complete the plan. But sometimes, I might forget to ask something important or miss a key detail.''}

As a result of the comprehensiveness of the inquiry process, most participants (16/22) found that solutions generated by CARE were more comprehensive, although not prominent in the quantitative analysis. A significant advantage of CARE, as noted by participants, was its ability to guide users through various dimensions of their task, ensuring that no critical aspects were overlooked. For instance, \textit{P16} shared: \textit{``A (CARE) really helped me think through every aspect of my travel plans. It wasn’t just a basic itinerary; it considered things like how much free time I would have, what activities were nearby, and even suggestions for restaurants based on my budget.''} 
Moreover, the structured organization of CARE’s generated solutions helped users follow and trust the planning process more easily, further enhancing their perception of the solution’s thoroughness. \textit{P19} noted that CARE’s step-by-step breakdown of tasks made the final solution more coherent: \textit{``A’s (CARE's) plan was laid out in clear steps, and it felt more complete because I could see how each decision built on the previous one. It wasn’t just a list of things to do; it was a logical progression that covered everything.''} 

In contrast, participants (7/22) noted that the baseline system often delivered less detailed plans, particularly due to its reactive nature. \textit{P12} described an instance where the baseline plan failed to include specific logistical considerations: \textit{``With B (Baseline), it felt like I was getting a quick summary rather than a full plan. It didn’t take into account things like transportation between activities, so I had to figure that out on my own.''} 

\subsubsection{Personalization Through Tailored Inquiry Results in More User-Specific Solutions}
CARE generates highly personalized plans by continuously adjusting its output based on users’ specific needs through dynamic questioning. 
Specifically, CARE’s inquiries focus on clarifying vague requests by asking about personalized preferences or needs, such as preferred activity types or priorities. This not only helps users articulate their preferences more clearly but also ensures that the system tailors its output accordingly. The flexibility and specificity of the generated solution made users feel that the system was responsive to their input, enhancing their satisfaction. \textit{P10} highlighted this by explaining, \textit{``A (CARE) kept asking questions that I hadn’t even thought about, like whether I wanted to prioritize certain skills or what time of day I preferred for training. This really helped make the final plan feel like it was made just for me.''} 

The baseline system often fell short in this area as it relied on users to provide all necessary input upfront without prompting further clarification. As a result, the solutions generated by the baseline system were often less personalized. \textit{P9} commented, \textit{``B (Baseline) didn’t ask any follow-up questions after I gave my preferences. It just generated a plan that was okay, but it didn’t really reflect everything I wanted.''} This feedback underscores how baseline’s limited engagement led to less tailored solutions, resulting in solutions felt generic. Even though occasionally the baseline system encouraged users to follow up with additional inputs, long-form answers could still lead to missing crucial personalization elements. For instance, \textit{P13} noted: \textit{``For my fitness training task, the plan from B (Baseline) didn’t match what I wanted at first. I had to follow up and ask for revisions, but even after updating, it still missed some details that A (CARE) had caught because A asked very specific questions from the start.''} This illustrates that even when users try to supplement their preferences with follow-up inquiries in the baseline system, the lack of thorough questioning from the outset could result in oversights.

CARE’s adaptability during the interaction process further enhanced its ability to personalize plans. \textit{P8} appreciated how the system dynamically adjusted based on real-time inputs: \textit{``With A (CARE), I could tweak my needs as we went along, and it would adjust the plan. That made me feel like I had control over the outcome and that the plan was really made for me.''} This iterative refinement process ensured that the final plan was more accurately tailored to the user’s evolving preferences, providing a clear advantage over the more static baseline system.


\subsubsection{CARE's Inquiry Process May Occasionally Cause Friction for Some Users.}

 While some participants (9/22) explicitly expressed that the CARE system’s structure made it better suited for guiding users through complex tasks, with \textit{P20} noting: \textit{``A (CARE) helps break down my larger tasks into smaller, more manageable parts, making it easier to navigate and ensure all details are covered''}, some participants (4/22) commented that the number of questions felt excessive at times. For instance, \textit{P13} noted: \textit{``A (CARE) sometimes provided new information, and while some of it was helpful, I also felt that when it kept asking about things I didn’t really care about, I wasn’t sure how to answer.''} This feedback suggests that users who prefer to resolve issues quickly may find the step-by-step inquiry process to be a source of mild friction.

\subsection{CARE's Collaborative Workspace Effectively Organizes User Needs, Generated Solutions, and Chat Interactions, thereby Reducing Cognitive Load}

One of the most frequently mentioned advantages of the CARE system during interviews was its well-organized UI. Many participants (14/22) noted that compared to the baseline, CARE's UI significantly improved their overall experience by reducing cognitive load and making interactions more intuitive. Specifically, users appreciated the structured layout, which clearly separated their needs from generated solutions and system interactions. This organization helped them stay focused on their tasks and reduced the effort needed to manage the flow of information during extended interactions. In the following sections, we will delve deeper into how specific features of CARE's UI, such as the \textit{Needs Panel} and \textit{Solution Panel}, contributed to these advantages.

\subsubsection{Panel Separation Improved Clarity and Reduced Cognitive Load}

The division of the \textit{Chat Panel}, \textit{Solution Panel}, and \textit{Needs Panel} within CARE significantly enhances clarity and minimizes users' cognitive load. Isolating the generated solution from the chat eliminates the need to sift through previous messages, while the distinct presentation of needs keeps them readily visible and adjustable. This design allows users to keep track of their evolving objectives without being overwhelmed by unrelated content.

Several participants appreciated how the clear separation of chat and solution output streamlined their interaction with the system. For example, \textit{P14} remarked: \textit{“It (CARE) asks questions on the right side and outputs what I need on the left, so I don’t need to search through the conversation to find the result.”} This clear division between the chat interactions and the solution outputs allowed users to locate relevant information effortlessly, minimizing the cognitive effort associated with scrolling back through chat histories.

The \textit{Needs Panel}, in particular, proved to be a critical feature in reducing mental load. It enabled users to track and update their requirements in real time without revisiting previous conversations. \textit{P8} noted, \textit{“A’s (CARE's) needs list is clearly displayed, making it easy to see and update my requirements.”} This visualization of needs provided users with a constant, structured overview, reducing the effort required to recall or modify their requests. Similarly, \textit{P21} highlighted how the separation of the \textit{Needs Panel} from the chat log made it easier to manage goals: \textit{“Having the needs separate from the chat makes it easier to verify what I’ve said and what I want.”} This dedicated space for evolving goals ensured that critical information was not overlooked, providing users with confidence that their inputs were being considered.

Moreover, this clear organization reduced the cognitive burden associated with planning tasks. \textit{P19} shared: \textit{“A’s (CARE's) table makes everything much clearer, so I don’t have to keep everything in my head or scroll back to find important details.”} By visually organizing task details in the \textit{Needs Panel}, CARE allows users to focus on decision-making rather than on recalling details, making the interaction less mentally demanding.

In contrast, the unstructured chat format of the baseline system led to frequent cognitive strain for some participants (6/22). The unstructured chat format made it difficult to locate and recall information, requiring constant scrolling. As \textit{P16} described: \textit{“With B (Baseline), I have to scroll up and down constantly to find the information.”} This lack of clear organization increased users’ mental load, making the process more tedious and hindering their ability to maintain an overview of their plans.

\subsubsection{User Needs Panel Simplified Interaction and Built Trust Through Transparent Solution Mapping}
The CARE system's \textit{Needs Panel} emerged as a key feature that simplified interaction and enhanced user trust by providing a clear, real-time display of user requirements and linking those needs transparently to the generated solutions. The \textit{Needs Panel} was recognized by several participants (10/22) for providing a real-time display of user requirements, enabling them to manage preferences efficiently throughout the interaction. For instance, \textit{P14} explained: \textit{``There is a Needs Panel where I can directly input my needs, and I don’t have to worry about the AI missing any of them.''}

This dynamic interaction was further supported by the real-time updating of needs, which allowed users to adapt their goals on the fly, improving the personalization of the outcomes. \textit{P17} highlighted this benefit: \textit{``A’s (CARE's) ability to update my needs in real-time led to more accurate and personalized plans. In contrast, B (Baseline) would often forget or shift focus after multiple conversations, making it harder to track everything.''} 

Moreover, CARE’s user needs reference mechanism further reinforced user trust by linking specific parts of the generated solutions back to the corresponding user needs, as shown by the linkage of the \textit{Needs Panel} and \textit{Solution Panel} in Figure~\ref{fig:ei_sys_fig}. Several participants (7/22) appreciated how this transparency reassured them that their input was being accurately incorporated. \textit{P21} commented, \textit{``The need references in A (CARE) are very useful because they show me which parts of the plan were based on my input, so I know exactly how my needs were considered.''} This clear mapping made it easier for users to understand and adjust the solutions, fostering a more engaged and confident interaction. As \textit{P14} noted, \textit{``With A (CARE), I could always see which parts of the generated solution were directly related to what I had asked for, which made it much easier to make adjustments.''} 

Participants consistently highlighted that this transparent connection between user needs and solution mapping increased their trust and sense of ownership. \textit{P8} observed, \textit{``Seeing how my requirements were implemented step-by-step in A (CARE) gave me a lot of confidence that the system was understanding and working with my needs accurately.''} In contrast, the baseline system, lacking a similar mechanism, often left users uncertain about how their input was utilized, leading to frustration. \textit{P22} expressed: \textit{``In B (Baseline), I wasn’t always sure if the system really understood my needs, since there was no way to see how my input was reflected in the final output.''}

Overall, the combination of CARE’s \textit{Needs Panel} and user needs reference features simplified the user experience and built trust by providing a transparent and tangible link between user input and system output. This transparency not only improved user confidence but also empowered them to make more informed and precise adjustments, resulting in a more reliable and user-centric planning process.

\subsubsection{Enhanced Readability of the Solution Panel through Structured and Rich-Text Formatting}
The structured presentation in CARE's \textit{Solution Panel}, using Markdown elements such as tables, rich text, and icons, was highlighted by participants as a key factor in enhancing readability.

Several users (6/22) appreciated the structured format, particularly in contrast to the unstructured chat format of the Baseline system. \textit{P2} noted, \textit{“A’s (CARE’s) tables, icons, and emojis made everything much more organized and readable. If it were only text, I might not continue the conversation after a few exchanges.”} This organized layout helped users understand and follow the system's generated plans with less cognitive effort. \textit{P16} similarly highlighted the benefits of the table-based layout: \textit{“A (CARE) puts everything into tables, which makes my schedule much clearer. With B (Baseline), I constantly have to scroll to find the information.”} This layout made navigating complex information more straightforward, reducing the need for excessive scrolling and minimizing cognitive overload.

In summary, the CARE system's organized UI reduces cognitive load by clearly separating user needs, solutions, and chats. Features like the \textit{Needs Panel} and \textit{Solution Panel} provide real-time clarity, helping users manage tasks efficiently and with less effort. This structured design enhances ease of use, trust, and reduces the strain of managing information compared to the baseline system.

\section{Discussion}

\subsection{Broader Implications for Collaborative Chat-Based Interfaces}
The findings from our study provide several important insights into designing more effective, collaborative chat-based systems, particularly for exploratory tasks. A key takeaway is the need to shift from reactive, dialogue-based interactions to a more proactive and structured approach, as demonstrated by the CARE system. CARE's ability to reduce cognitive load and support more meaningful engagement with users stems from its structured interface and adaptive collaboration, setting it apart from traditional, free-flowing conversational agents.

One of the broader design principles emerging from our study is the separation of conversational inputs from task management. Many LLM-based systems rely heavily on an undifferentiated, continuous dialogue format, which can overwhelm users and make it difficult to organize and track complex tasks. In contrast, CARE’s dual-panel approach—distinguishing between conversational inputs (\textit{Chat Panel}) and structured outputs (\textit{Solution Panel})—proves highly effective in managing information. By separating these components, users can focus on task progression without being bogged down by unnecessary dialogue or clutter. A dedicated \textit{Needs Panel}, designed to ensure traceability and transparency between user inputs and system outputs, enhances user confidence and trust in the system. This suggests that future chat-based systems could benefit from integrating multimodal interaction spaces that visually differentiate between ongoing conversation, system outputs, and task organization. Such interfaces could vastly improve user experience by mitigating information overload and making it easier to manage evolving, multi-step processes.

Furthermore, CARE’s design prioritizes adaptive collaboration. Unlike conventional systems that passively respond to user inputs, CARE actively inquires about user needs, offers suggestions, and encourages exploration. This interaction model fosters creativity and supports personalization, particularly in open-ended tasks where users might not have well-defined goals. By dynamically adjusting to users' evolving needs and preferences, CARE transforms the interaction from one of mere conversational efficiency to an ongoing, exploratory dialogue. This agent-driven approach could be particularly valuable in domains such as complex decision-making, creative processes, or long-term project management. For the broader HCI community, the CARE system exemplifies how conversational agents can evolve from static responders into active collaborators, promoting more exploratory and creative problem-solving.

Additionally, our findings reveal that the CARE system helps uncover implicit user needs by inspiring users to think beyond their initial queries. Rather than just providing direct answers, CARE prompts users to consider alternative perspectives and explore unanticipated paths, creating a more enriching interaction. This design feature could inform future work on intelligent agents, which need to balance guidance and autonomy. Systems that empower users to extend their thinking without overwhelming them represent a significant opportunity for future conversational interfaces, particularly in areas like education, content creation, and collaborative work environments. The broader implication for HCI is a shift towards systems that don’t just provide information but actively support users in navigating uncertainty, promoting discovery, and enhancing creativity.

\subsection{Limitations and Future Work}

While our study demonstrates CARE's potential in improving exploratory tasks, it has limitations that future research should address. CARE's response latency, inherent to its multi-agent design, poses a challenge. While the system excelled in reducing cognitive load and enhancing personalization, the added response time could affect user satisfaction. Advancements in LLM technology will help reduce these delays.
Another limitation is the relatively small and homogeneous participant pool, which may affect the generalizability of our findings. Future studies should recruit more diverse participants to capture a broader range of user needs, preferences, and interaction patterns.
Lastly, our study used GPT-4o, one of the most advanced LLMs at the time. Future research should investigate whether the findings generalize across other LLM implementations, especially as new, more efficient models emerge. Exploring alternative interaction modalities like voice or gesture could also further enhance the system’s capabilities.

\section{Conclusion}
In this paper, we introduced CARE, a collaborative chat-based interface designed to enhance personalized exploratory tasks by addressing key limitations in existing LLM-based systems. Motivated by challenges such as the lack of sustained contextual understanding and personalization in current LLM-based chatbots, we developed a system that reimagines user-LLM interaction through structured and adaptive collaboration. Our design goals focused on improving contextual understanding, facilitating personalization, inspiring user exploration, and reducing cognitive load.
Through a user study comparing CARE with a baseline LLM chatbot with 22 participants, the results demonstrated that CARE outperformed the baseline system in several key areas, notably in inspiring user exploration, personalizing responses, and organizing information effectively. These findings have broader implications for the design of future chat-based interfaces. By empowering users to navigate ambiguity and supporting them in complex tasks, systems like CARE have the potential to transform how users engage with conversational agents, making them not just reactive tools, but proactive partners in problem-solving and discovery.


\bibliographystyle{ACM-Reference-Format}
\bibliography{ref}

\appendix
\section{Appendix: User Study Materials}
\subsection{Follow-up Interview} \label{sec:appendix_interview}
During the follow-up interview, we asked following questions to the participants to gather additional qualitative insights, particularly focusing on any notable highlights or frustrations with each system.
\begin{itemize}
    \item Overall, which system do you prefer, Chatbot A or Chatbot B?
    \item Which interaction style do you prefer, Chatbot A or Chatbot B? Please describe in detail.
    \item Can you describe in detail which features or functions of the system were most helpful to you? Please provide examples.
    \item How well do you think the system's responses matched your needs? If they didn’t match, please describe what information or personalization you think the system overlooked.
    \item Did the system provide you with new inspiration or directions for your thinking during the exploratory tasks? Please explain in detail how you were inspired.
    \item What was your experience with the system? Were there any aspects that impressed you, or areas where you felt improvement is needed?
    \item Did you encounter any confusion or inconvenience while using the system? Please describe in detail.
    \item What do you think of the UI design?

\end{itemize}
\subsection{User Experiment Registration Questionnaire} \label{sec:appendix_survey1}
This questionnaire was used to collect basic demographic information from participants to assist in analyzing the data from this experiment. We greatly appreciate the participants' involvement in this study.

\begin{enumerate}
    \item \textbf{Your ID (e.g., P1):} \\
    \underline{\hspace{10cm}}
    
    \item \textbf{Your gender:} \\
    (Please select one)
    \begin{itemize}
        \item  Male
        \item  Female
        \item  Other (please specify) \underline{\hspace{3cm}}
        \item  Prefer not to disclose
    \end{itemize}
    
    \item \textbf{Your field of study:} \\
    (Please select one)
    \begin{itemize}
        \item Computer Science
        \item Engineering (e.g., Electrical, Mechanical)
        \item Natural Sciences (e.g., Physics, Chemistry)
        \item Social Sciences (e.g., Psychology, Economics)
        \item Arts
        \item Other (please specify) \underline{\hspace{3cm}}
    \end{itemize}
    
    \item \textbf{Your education level:} \\
    (Please select one)
    \begin{itemize}
        \item Undergraduate student
        \item Master's student
        \item PhD student
        \item PhD or above
    \end{itemize}

    \item \textbf{How familiar are you with conversational AI systems (e.g., ChatGPT, Kimi, Wenxin Yiyan, Tongyi Qianwen, etc.)?} \\
    (Please select one)
    \begin{itemize}
        \item Heard about it, but never used
        \item Used it, less than once per week
        \item Used it, 1-7 times per week
        \item Used it, multiple times per day
    \end{itemize}
\end{enumerate}

\subsection{Post-task User Interaction Experience with Conversational AI Questionnaire} \label{sec:appendix_survey2}

This questionnaire was used to gather feedback on participants' experience interacting with a conversational AI system. The feedback helps to improve our understanding of user experience with the system. We highly appreciate the participants' engagement in this experiment.

\begin{enumerate}
    \item \textbf{Your ID and the chatbot system you used (e.g., P1\_A):} \\
    \underline{\hspace{10cm}}

    \item \textbf{I enjoyed the way I interacted with the system.} \\
    (Please rate on a scale of 1 to 5, where 1 is strongly disagree and 5 is strongly agree)
    \begin{itemize}
        \item 1 (strongly disagree)
        \item 2 (disagree)
        \item 3 (neutral)
        \item 4 (agree)
        \item 5 (strongly agree)
    \end{itemize}

    \item \textbf{The system’s UI presentation helped me better organize complex tasks and reduced my cognitive load.} \\
    (Please rate on a scale of 1 to 5, where 1 is strongly disagree and 5 is strongly agree)
    \begin{itemize}
        \item 1 (strongly disagree)
        \item 2 (disagree)
        \item 3 (neutral)
        \item 4 (agree)
        \item 5 (strongly agree)
    \end{itemize}

    \item \textbf{The interaction with the system inspired me to consider new aspects of exploratory tasks.} \\
    (Please rate on a scale of 1 to 5, where 1 is strongly disagree and 5 is strongly agree)
    \begin{itemize}
        \item 1 (strongly disagree)
        \item 2 (disagree)
        \item 3 (neutral)
        \item 4 (agree)
        \item 5 (strongly agree)
    \end{itemize}

    \item \textbf{The answers provided by the system felt comprehensive and sufficient.} \\
    (Please rate on a scale of 1 to 5, where 1 is strongly disagree and 5 is strongly agree)
    \begin{itemize}
        \item 1 (strongly disagree)
        \item 2 (disagree)
        \item 3 (neutral)
        \item 4 (agree)
        \item 5 (strongly agree)
    \end{itemize}

    \item \textbf{The answers provided by the system matched my personal needs.} \\
    (Please rate on a scale of 1 to 5, where 1 is strongly disagree and 5 is strongly agree)
    \begin{itemize}
        \item 1 (strongly disagree)
        \item 2 (disagree)
        \item 3 (neutral)
        \item 4 (agree)
        \item 5 (strongly agree)
    \end{itemize}

\end{enumerate}

\section{Agent Prompts}\label{sec:appendix_prompt}

In this appendix, we provide an overview of the prompts used by the five LLM-based agents in our system, along with a summary of the user survey employed for evaluation. The agents—\textit{Inquiry Agent}, \textit{Milestone Agent}, \textit{Needs Discovery Agent}, \textit{Ranking Agent}, and \textit{Solution Craft Agent}—each have specific prompts that guide their interactions and decision-making processes. These prompts are designed to help the agents autonomously assist users in identifying their needs and generating solutions. The following sections detail the prompts for each agent.

\subsection{Team Introduction Prompt}
We use a team introduction to provide each agent with information about the team. This introduction is written at the very beginning of each agent's section. The introduction is in Table~\ref{tab:intro}.

\begin{table*}[!ht]
    \setlength{\abovecaptionskip}{0.1cm}
    \setlength{\belowcaptionskip}{-0.5cm}
    \centering
    
    \footnotesize
    \begin{tabular}{p{15cm}}
    \toprule
        \# Team Introduction \\
        You are part of a versatile team that specializes in solving a wide variety of user needs. \\
         \\
        \#\# Team Member Introduction \\
        Your team includes: \\
        1. Inquiry Agent: Responsible for direct communication with users, including asking for basic information, understanding user preferences and needs, and collecting user feedback on solutions. \\
        2. Milestone Agent: Responsible for determining the next major direction for the current task. \\
        3. User Needs Discovery Agent: Responsible for identifying the user's needs related to the current task. \\
        4. Planning Agent: Responsible for creating personalized solutions based on the user needs uncovered by the team. \\
        5. Ranking Agent: Responsible for grouping and then ordering the clarification questions.  \\
         \\
        \#\# Team Goal \\
        The goal of your team is to solve various user problems and provide personalized solutions. To provide these personalized solutions, the team will first explore the user's preferences and needs before presenting a solution. In addition to the needs explicitly stated by the user, the team hypothesizes implicit user needs and verifies these through communication with the user. \\
        \#\#\# Personalized Solutions \\
        Your team uses a tool called User Needs Memo to store possible user needs. The User Needs Memo is visible and editable by the user. Below is an introduction to the format of the User Needs Memo: \\
        \#\#\#\# User Needs Memo \\
        The User Needs Memo is a JSON-formatted dictionary where each key represents a unique\_id, which is automatically generated by the system. Team members can use this unique\_id to retrieve the corresponding user need. The value associated with the key represents a Need Slot. \\
        \#\#\#\#\# unique\_id \\
        The unique\_id is a unique identifier generated by the uuid library. \\
        \#\#\#\#\# Need Slot \\
        A Need Slot is a dictionary containing two keys: \\
        \begin{verbatim}
{
    "need": "The detailed description of need",
    "Clarify": true/false,
}
        \end{verbatim}
        1. need: If Clarify=true, it indicates the specific description of the need. If Clarify=false, it represents a question to ask the user in order to clarify and obtain the final description of the user's need. \\
        2. Clarify: Indicates whether it is necessary to ask the user if they want this need. \\
         \\
        \#\#\#\#\# User Need Categories \\
        User needs can be divided into three categories: \\
        1. **Explicit Needs**: Needs explicitly stated by the user. These are needs that the user has clearly expressed. These needs must be fully collected. If these needs are not met, it will cause great dissatisfaction, but meeting them will not increase satisfaction. The keys in the Need Slot should be set as follows: \\
            - Clarify=false \\
        2. **Implicit Needs**: Needs not explicitly stated by the user but of which the user is **aware**. These needs should be collected as fully as possible. These requirements directly affect satisfaction. Meeting them increases satisfaction, while not meeting them leads to dissatisfaction. The keys in the Need Slot should be set as follows: \\
            - Clarify=true \\
        3. **Latent Needs**: Needs that the user is **unaware** of, but which do exist. These requirements exceed customer expectations. Meeting them brings great satisfaction, but not meeting them does not cause dissatisfaction. To better satisfy these needs, the team needs to continuously explore the user's unrecognized needs. The keys in the Need Slot should be set as follows: \\
            - Clarify=true \\
         \\
        \#\#\#\#\# Format Example \\
        \begin{verbatim}
{
    "0": {
        "need": "The travel destination is Tokyo.",
        "Clarify": false,
    },
    "1": {
        "need": "What type of accommodation does the user prefer?",
        "Clarify": true,
    },
    ...
}
        \end{verbatim}
        
        The 0, 1 are the id, which is an automatically assigned incremental ID by the system, and you cannot modify it. \\
        \#\# Language use \\
        At the beginning of conversation, you should decide the language used to chat with user. 
        - **All of your response must be in English!** \\
        \\
        \\
    \bottomrule
    \end{tabular}
    \caption{The prompt of Team Introduction}
    \label{tab:intro}
\end{table*}

\subsection{Inquiry Agent Prompt}
The primary role of the \textit{Inquiry Agent} is to inquire about and clarify user needs by simplifying questions, offering options, and providing explanations, thereby reducing cognitive load and improving the interaction process. We show it's prompt in Table~\ref{tab:inquiry}.

\begin{table*}[htbp]
    \setlength{\abovecaptionskip}{0.1cm}
    \setlength{\belowcaptionskip}{-0.5cm}
    \centering
    \footnotesize
    \begin{tabular}{p{15cm}}
    \toprule
    You are now serving as the \texttt{\`}Inquiry-Agent\texttt{\`} and working with an outstanding team. Below is your team introduction: \\
    \{team\_intro\} \\
     \\
    Here is your role introduction and work content: \\
     \\
    \#\# Role Introduction \\
    As the \texttt{\`}Inquiry-Agent\texttt{\`}, you are the only member of the team capable of communicating with the user. When interacting with the user, you must ensure a friendly and approachable tone. While communicating, you should continuously gather the user's requirements. \\
     \\
    \#\# Work Content \\
    1. At the beginning, the user will provide you with a query. You need to pass the user's initial query exactly as it is to the Milestone-Agent (Note: You do not need to call any functions for this step). At the end, you should generate \texttt{\`}[BeginMilestone]\texttt{\`}. Here is a simple example: \\
    \begin{verbatim}
```markdown
    some text to tell Milestone-Agent what user query is... (You must write the detail of user query in the text)
    [BeginMilestone]
```
    \end{verbatim}
    2. The \texttt{\`}Ranking-Agent\texttt{\`} will give you some group questions. Then, you need to ask the user questions follow the order that QuestionRefine-Agent gives you to understand their actual needs. Before asking the questions, you should think step by step: \\
    1. Before asking questions from a group, you can ask the user if they have any needs in that area. If the user feels that there are no needs, you can skip all the questions in that group. If the user thinks the group content is necessary, you can proceed with asking questions. \\
    2. Only ask questions from one group at a time. If there are too many questions in one group, break them up, asking **3~4 questions** at a time until all the questions in the group are covered. \\
        - When asking questions, you need to simplify them to ensure the user can understand. \\
        - For some questions, you need to provide **default options**. For example: "What kind of animal do you like? Cat or dog?" \\
     \\
    3. After the user answers, you need to fill in the \texttt{\`}Need Slots\texttt{\`} requiring clarification by calling the \texttt{\`}fill\_need\_slot\texttt{\`} function. For the \texttt{\`}need\texttt{\`} parameter, you should be as detailed as possible. For example, if the requirement is the user's address, you should write: The user lives in China. Rather than just writing China. \\
    4. **At the end of your questions, you MUST generate: \texttt{\`}[Inquiry]\texttt{\`}.**  \\
    5. Here is a simple example for asking user questions: \\
    \begin{verbatim}
```markdown
    some polite and encouraging text to user...
    1. Question 1: ...
    2. Question 2: ...
    ...
    n. Question n: ...
    [Inquiry]
```
    \end{verbatim}
    4. After all the questions have been asked, you need to inform the Milestone-Agent to get next inquiry focus. At the end, you should generate \texttt{\`}[BeginMilestone]\texttt{\`} \\
     \\
    5. After the SolutionCraft-Agent has formulated the Solution, he will inform you, and you need to send a message to the user to tell them the solution is ready. But you do not need to tell the user the specific content of the solution. Just remind the user to check the solution. \\
     \\
    6. After user has check the solution, he/she will review it and provide feedback. You need to organize the user's feedback and convey it to the Milestone-Agent. Afterward, other Agents will write any new needs and potential needs raised by the user into the \texttt{\`}User Needs Memo\texttt{\`}. \\
     \\
    7. Special reminder: If the user explicitly states that they don't want to answer questions and want to see the solution immediately, you should stop asking questions right away. Notify the Milestone-Agent that the user wants to generate an answer immediately. If the user has provided any feedback, include that feedback when informing the Milestone-Agent. \\
     \\
    8. If the user informs you that they have manually updated their requirements, you should immediately notify the Milestone-Agent about this update. Inform them that the user has updated their own requirements. \\
     \\
    \# Attention  \\
    1. You can only call functions: \texttt{\`}[fill\_need\_slot]\texttt{\`}. YOU CANNOT CALL ANY OTHER FUNCTION NAME. It will cause serious disaster. \\
    \\
    \\
    \bottomrule
    \end{tabular}
    \caption{The prompt of Inquiry Agent.}
    \label{tab:inquiry}
\end{table*}

\subsection{Milestone Agent Prompt}
The primary role of the \textit{Milestone Agent} is to coordinate the process by setting actionable steps based on user needs to ensure the solution process progresses smoothly. We show it's prompt in Table~\ref{tab:milestone}.

\begin{table*}[htbp]
    \setlength{\abovecaptionskip}{0.1cm}
    \setlength{\belowcaptionskip}{-0.5cm}
    \centering
    \footnotesize
    \begin{tabular}{p{15cm}}
    \toprule
    You are now serving as a \texttt{\`}Milestone-Agent\texttt{\`} and working with an excellent team. Here is an introduction to your team:  
    \{team\_intro\} \\
     \\
    Below is an introduction to your role and responsibilities: \\
     \\
    \#\# Role Introduction \\
     \\
    As a \texttt{\`}Milestone-Agent\texttt{\`}, you have two responsibilities: \\
     \\
    1. When the user believes the solution needs improvement, or if you think more specific requirements from the user are needed, you need to think about the next milestone for the team based on user queries, the current recorded user needs, previously established milestones, and any user feedback (if available). \\
     \\
    2. When you believe that the current collected requirements are sufficient to formulate or modify the solution, you need to notify the \texttt{\`}SolutionCraft-Agent\texttt{\`} to begin developing the solution. \\
     \\
    \#\# Milestone Introduction \\
     \\
    - A milestone refers to a key area that the team needs to prioritize. It mainly includes the following aspects: \\
      1. Collecting the user's basic personal information (Note: Only collect information relevant to solving the task; avoid collecting unnecessary information that infringes on user privacy). \\
      2. Planning sub-tasks for the main user's query. \\
    - Milestones must be specific goals and not overly vague. For example, it cannot be: Satisfy user feedback. \\
    - You **cannot set milestones that have already been established**, as this may lead to user dissatisfaction. \\
     \\
    \#\# Responsibilities \\
     \\
    In each round, you need to use the \texttt{\`}get\_all\_needs\texttt{\`} function to retrieve the recorded user needs, which include both \texttt{\`}User Wants Needs\texttt{\`} and \texttt{\`}User do not want to answer needs\texttt{\`}. You can not bulid a solution based on \texttt{\`}User do not want to answer needs\texttt{\`}.  \\
     \\
    You should design milestones based on the user's current feedback and recorded needs. Then, call the \texttt{\`}load\_solution\texttt{\`} function to get the current solution [Note: \texttt{\`}load\_solution\texttt{\`} may return empty, as solutions may not have been developed yet]. When setting the next milestone, you need to refer to the existing user needs and already established solutions, and consider the user's query/feedback. You must follow these guidelines: \\
     \\
    1. If the \texttt{\`}User Needs Memo\texttt{\`} is empty, the first milestone should be: Collect detailed basic user needs required to complete the task. \\
     \\
    2. If the \texttt{\`}User Needs Memo\texttt{\`} is not empty, and you believe the current needs are insufficient to complete the task, you need to determine the next milestone based on the currently recorded user needs and user feedback. After generating the next milestone, you need to clearly inform the \texttt{\`}UserNeedsDiscovery-Agent\texttt{\`} about the next milestone and the user's query/feedback. Additionally, you should provide an explanation of why this milestone is being focused on. Finally, generate \texttt{\`}[MilestoneEnd]\texttt{\`}. For example: \\
    \begin{verbatim}
```
Next milestone:....
    - Explanation:...
User query/feedback:...
[MilestoneEnd]
```
    \end{verbatim}
    3. If the \texttt{\`}User Needs Memo\texttt{\`} is not empty, and you believe the current recorded needs are sufficient to address the user's query or the user want to directly begin planning, you need to notify the \texttt{\`}SolutionCraft-Agent\texttt{\`} to start generating a solution based on the \texttt{\`}User Needs Memo\texttt{\`}. Besides, you should tell the \texttt{\`}SolutionCraft-Agent\texttt{\`} the user's query/feedback. Finally, generate \texttt{\`}[BeginPlan]\texttt{\`}. At this point, you do not need to set a milestone. For example: \\
    \begin{verbatim}
```
User query/feedback:...
[BeginPlan]
```
    \end{verbatim}
    4. If the Inquiry-Agent notifies you that the user has manually updated their requirements, immediately notify the Planning Module to begin planning. Generate \texttt{\`}[BeginPlan]\texttt{\`} and include any information about the user's updates. For example: \\
    \begin{verbatim}
```
User has updated their requirements by themselves.
[BeginPlan]
```
    \end{verbatim}
    \textbf{CONTINUE ON THE NEXT PAGE} \\
    \\
    \bottomrule
    \end{tabular}
    \caption{The prompt of Milestone Agent.}
    \label{tab:milestone}
\end{table*}

\begin{table*}[htbp]
    \setlength{\abovecaptionskip}{0.1cm}
    \setlength{\belowcaptionskip}{-0.5cm}
    \centering
    \footnotesize
    \begin{tabular}{p{15cm}}
    \toprule
    \#\# Guidelines for Creating Effective Milestones \\
     \\
    When creating milestones, follow these guidelines to ensure they are specific, actionable, and valuable: \\
     \\
    1. Be specific and measurable: Each milestone should have a clear, concrete outcome that can be easily verified. \\
     \\
    2. Align with user goals: Ensure that each milestone directly contributes to addressing the user's main query or problem. \\
     \\
    3. Prioritize based on importance: Focus on the most critical aspects of the task first. \\
     \\
    4. Break down complex tasks: If a task seems too large, break it into smaller, manageable milestones. \\
     \\
    5. Consider dependencies: Think about the logical order of steps and any prerequisites for each milestone. \\
     \\
    6. Adaptable: Be prepared to adjust milestones based on new information or changing user needs. \\
     \\
    7. User-centric: Frame milestones in terms of user benefits or progress towards their goal. \\
     \\
    8. Avoid redundancy: Ensure each new milestone adds unique value and doesn't overlap with previous ones.
     \\
    9. Balance detail and flexibility: Provide enough detail for clarity, but allow room for the team to determine the best approach. \\
     \\
    \#\# Examples of Good Milestones \\
     \\
    - "Identify the top 3 pain points in the user's current workflow" \\
    - "Define the core features of the proposed solution based on user needs" \\
    - "Create a prioritized list of user requirements for the new system" \\
    - "Develop a high-level architecture diagram for the proposed solution" \\
    - "Outline the key performance indicators (KPIs) for measuring the solution's success" \\
     \\
    Remember, effective milestones guide the team towards a clear goal while allowing for discovery and adaptation along the way. \\
     \\
    \#\# Notes \\
     \\
    1. If the \texttt{\`}User Needs Memo\texttt{\`} contains user information that is uncertain, you should not proceed with setting a milestone. This is because the information is not clear enough for the user and needs to be clarified by the team's SolutionCraft-Agent. \\
     \\
    2. When you are not calling functions, you **must generate \texttt{\`}[BeginPlan]\texttt{\`} or \texttt{\`}[MilestoneEnd]\texttt{\`}**. If you are calling \texttt{\`}get\_all\_needs\texttt{\`} or \texttt{\`}load\_solution\texttt{\`}, you should not generate these markers. \\
     \\
    \#\#\# How to Determine if Current Recorded Needs Can Address the User's Query \\
     \\
    1. If the user has not provided feedback, but the current recorded needs are insufficient to complete the task, you need to determine the next milestone based on the currently recorded user needs and user feedback. \\
    \\
    \\
    \bottomrule
    \end{tabular}
    \caption{The prompt of Milestone Agent.}
    \label{}
\end{table*}

\subsection{Needs Discovery Agent Prompt}
The primary role of the \textit{Needs Discovery Agent} is to identify both explicit and implicit user needs for comprehensive documentation. We show it's prompt in Table~\ref{tab:discovery}.

\begin{table*}[htbp]
    \setlength{\abovecaptionskip}{0.1cm}
    \setlength{\belowcaptionskip}{-0.5cm}
    \centering
    \footnotesize
    \begin{tabular}{p{15cm}}
    \toprule
    You are now serving as the \texttt{\`}NeedsDiscovery-Agent\texttt{\`} and working with an outstanding team. Below is your team introduction: \\
    \{team\_intro\} \\
     \\
    Here is your role introduction and work content: \\
     \\
    \#\# Role Introduction \\
    As the \texttt{\`}NeedsDiscovery-Agent\texttt{\`}, you are responsible for identifying users' needs according to the theory, with a focus on uncovering implicit and latent needs that align with the current milestone. \\
     \\
    \#\# Workflow \\
    The \texttt{\`}Milestone-Agent\texttt{\`} will determine the next Milestone and inform you of the user's query/feedback. After understanding the user's requirements and the current milestones, you need to identify the user's needs and add them to the \texttt{\`}User Needs Memo\texttt{\`}. To achieve this goal, you need to **think step by step** and complete the following three steps: \\
     \\
    1. Call the \texttt{\`}get\_all\_needs\texttt{\`} function to retrieve all the existing user needs, including \texttt{\`}User Wants Needs\texttt{\`} and \texttt{\`}User Not Answered Needs\texttt{\`}. **You can not propose a new question, including in \texttt{\`}User Not Answered Needs\texttt{\`}**, otherwise, it will cause user dissatisfaction. \\
     \\
    2. Extract the explicit needs expressed by the user in the query. Let's think step by step: \\
        1. Do not extract needs that **exist in \texttt{\`}User Needs Memo\texttt{\`}** again, you should check it first.  \\
        2. All explicitly extracted requirements must be clearly stated by the user. For example, if the user says: "I want to travel to the US in the summer," you need to extract two explicit needs: \\
            1. Travel destination is the US. \\
            2. Travel date is in the summer. \\
            3. After extraction, you need to call the \texttt{\`}add\_need\_slot\texttt{\`} function, set \texttt{\`}need\texttt{\`} to the extracted user need, \texttt{\`}user\_want\texttt{\`} to \texttt{\`}true\texttt{\`}, and \texttt{\`}Clarify\texttt{\`} to \texttt{\`}false\texttt{\`}. You must ensure that all these needs are extracted since they are the user's basic needs. If these needs are not met, the user will be very dissatisfied. \\
     \\
    3. Identify the User's Implicit and Latent Needs that are **not mentioned in the \texttt{\`}User Needs Memo\texttt{\`}**. Focus on needs that align with the current milestone and contribute to its completion. Consider the following guidelines: \\
 \\
    - Analyze the current milestone and break it down into key components or aspects that need to be addressed. \\
    - For each component, brainstorm potential implicit or latent needs that could be relevant. \\
    - Consider the user's context, background, and any information provided in the \texttt{\`}User Needs Memo\texttt{\`}. \\
    - Think about potential challenges, preferences, or constraints the user might have related to the milestone. \\
    - Anticipate future needs or potential issues that might arise as the user progresses towards their goal. \\
 \\
    Examples of milestone-focused questions: \\
 \\
    - For the milestone "Identify the top 3 pain points in the user's current workflow": \\
        1. What specific tasks in the user's workflow are most time-consuming? \\
        2. Are there any recurring issues or bottlenecks in the current process? \\
        3. How does the user currently measure productivity or efficiency? \\
        4. What tools or systems is the user currently using, and what are their limitations? \\
        5. How do these pain points affect other team members or departments? \\
 \\
    - For the milestone "Define the core features of the proposed solution based on user needs": \\
        1. What are the user's primary goals when using the solution? \\
        2. How does the user envision interacting with the solution on a daily basis? \\
        3. Are there any industry-specific requirements or standards that need to be considered? \\
        4. What level of technical Agentise does the user have? \\
        5. How important is scalability or future expansion of the solution to the user? \\
 \\
    - For the milestone "Create a prioritized list of user requirements for the new system": \\
        1. What are the must-have features versus nice-to-have features for the user? \\
        2. How does the user define success for this new system? \\
        3. Are there any budget or time constraints that might affect prioritization? \\
        4. How do the requirements align with the user's long-term business goals? \\
        5. Are there any regulatory or compliance requirements that need to be considered? \\
 \\
    Once these needs are identified, use \texttt{\`}add\_need\_slot\texttt{\`} to update the \texttt{\`}User Needs Memo\texttt{\`}. Set \texttt{\`}need\texttt{\`} to the user's implicit need phrased as a question, set \texttt{\`}user\_want\texttt{\`} to \texttt{\`}null\texttt{\`}, and \texttt{\`}Clarify\texttt{\`} to \texttt{\`}true\texttt{\`}. \\
    \\
    \textbf{CONTINUE ON THE NEXT PAGE} \\
    \\
    \\
    \bottomrule
    \end{tabular}
    \caption{The prompt of Needs Discovery Agent.}
    \label{tab:discovery}
\end{table*}

\begin{table*}[htbp]
    \setlength{\abovecaptionskip}{0.1cm}
    \setlength{\belowcaptionskip}{-0.5cm}
    \centering
    \footnotesize
    \begin{tabular}{p{15cm}}
    \toprule
    \#\# Guidelines for Effective Need Discovery \\
     \\
    1. Be comprehensive: Consider all aspects of the milestone and how they relate to the user's overall goal. \\
    2. Think long-term: Anticipate future needs or challenges that may not be immediately apparent. \\
    3. Consider context: Take into account the user's industry, role, and specific circumstances. \\
    4. Be specific: Frame questions in a way that encourages detailed, actionable responses. \\
    5. Prioritize value: Focus on needs that, if addressed, would provide the most significant benefit to the user. \\
    6. Avoid assumptions: Don't assume you know the user's preferences or constraints without evidence. \\
    7. Consider interdependencies: Think about how different needs might interact or affect each other. \\
    8. Be user-centric: Always frame needs and questions from the user's perspective. \\
    9. Avoid direct translation: Do not simply rephrase the milestone explanation as needs. Instead, think critically about what underlying needs the milestone implies. \\
    10. Focus on actionable insights: Generate needs that will lead to specific, actionable information rather than general confirmations of the milestone itself. \\
     \\
    \# Attention \\
    1. You MUST call \texttt{\`}add\_need\_slot\texttt{\`} when you generate the needs. \\
    2. You can only call functions: \texttt{\`}[add\_need\_slot, get\_all\_needs]\texttt{\`}. YOU CANNOT CALL ANY OTHER FUNCTION NAME. It will cause a serious disaster. \\
    3. Only after adding all needs to \texttt{\`}User Needs Memo\texttt{\`}, you can generate \texttt{\`}[DISCOVEREND]\texttt{\`}. \\
    4. Do not directly translate milestone explanations into needs. Instead, think critically about what specific information or insights would be most valuable to achieve the milestone. \\
    \\
    \\
    \bottomrule
    \end{tabular}
    \caption{The prompt of Needs Discovery Agent.}
    \label{}
\end{table*}

\subsection{Ranking Agent Prompt}
The primary role of the \textit{Ranking Agent} is to organize and prioritize questions to streamline user interactions. We show the it's prompt in Table~\ref{tab:ranking}.
\begin{table*}[htbp]
    \setlength{\abovecaptionskip}{0.1cm}
    \setlength{\belowcaptionskip}{-0.5cm}
    \centering
    \footnotesize
    \begin{tabular}{p{15cm}}
    \toprule
    You are now serving as the \texttt{\`}Ranking-Agent\texttt{\`} and working with an outstanding team. Below is your team introduction:  
    \{team\_intro\} \\
     \\
    Here is your role introduction and work content: \\
     \\
    \#\# Role Introduction \\
    As the \texttt{\`}Ranking-Agent\texttt{\`}, you are responsible for grouping and then ordering the questions that need clarification, as identified by the \texttt{\`}NeedsDiscovery-Agent\texttt{\`}. \\
     \\
    \#\#\# Workflow \\
    You need to think step by step and give the explanation: \\
    1. First, you need to call the \texttt{\`}get\_clarify\_needs\texttt{\`} function to retrieve all \texttt{\`}Need Slots\texttt{\`} in the \texttt{\`}User Needs Memo\texttt{\`} that require clarification. \\
    2. Group all the questions that need clarification. \\
    3. While sorting the questions within each group, you also need to sort the order of the groups. \\
    4. Finally, generate a json-formatted text that follows the format of the example below: \\
    \begin{verbatim}
```json
{{
    "Topic 1": {{
        "question-1": {{
            "need_id": "The id of user need.",
            "need": "the clarification question."
        }},
        "question-2": {{
            "need_id": "The id of user need.",
            "need": "the clarification question."
        }},
        ...
    }},
    "Topic 2": {{
        "question-1": {{
            "need_id": "The id of user need.",
            "need": "the clarification question."
        }},
        ...
    }}

}}
""
...
```
    \end{verbatim}
    The principles for grouping are as follows: \\
    \#\#\# Grouping Principles \\
    1. The span of questions within a group should not be too broad, ensuring that the user feels they can answer the questions continuously and smoothly. \\
    2. The questions within a group must have a central theme, and all questions must revolve around this theme. \\
    3. Questions within a group should not affect each other; the answer to one question should not influence the answers to other questions. \\
     \\
    The principles for ordering are as follows: \\
    \#\#\# Ordering Principles \\
    1. Ordering questions within a group: Since the questions within a group are focused on a single theme, the order of the questions should ensure a progression from easy to difficult, providing a good user experience during answering. \\
    2. Ordering of question groups: There should be a logical sequence between groups, ensuring a progression from simple to complex. \\
    \\
    \\
    \bottomrule
    \end{tabular}
    \caption{The prompt of Ranking Agent.}
    \label{tab:ranking}
\end{table*}

\subsection{Solution Craft Agent}
The primary role of the \textit{Solution Craft Agent} is to generate personalized, actionable solutions based on the collected user input. We show it's prompt in Table~\ref{tab:solution}.
\begin{table*}[htbp]
    \setlength{\abovecaptionskip}{0.1cm}
    \setlength{\belowcaptionskip}{-0.5cm}
    \centering
    \footnotesize
    \begin{tabular}{p{15cm}}
    \toprule
    You are the \texttt{\`}SolutionCraft-Agent\texttt{\`}, a crucial member of an elite team developing personalized solutions for users. Your role is to create comprehensive, tailored plans that address all user needs effectively. \\
 \\
    \#\# Solution Development Process \\
     \\
    1. **Analyze User Needs**  \\
       - Retrieve current user requirements using \texttt{\`}get\_user\_want\_needs\texttt{\`}. \\
       - Compare with previous needs, identifying new or changed requirements. \\
       - Assign unique IDs to each need (e.g., \texttt{\`}Need ID: 001\texttt{\`}, \texttt{\`}Need ID: 002\texttt{\`}). And write the explanation in a \texttt{\`}>\texttt{\`} block. The IDs you reference must exist in the User Needs Memo, do not fabricate them. Otherwise, the user will be very confused and annoyed. \\
     \\
    2. **Develop Personalized Solution**  \\
       - Address each user need comprehensively and systematically. \\
       - Create specific, actionable plans for every aspect of the solution. \\
       - Provide clear explanations linking solutions to user requirements. \\
       - Offer reasonable suggestions for any omitted information based on context. \\
     \\
    3. **Implement Personalization Strategies**  \\
       - Analyze the user's situation, preferences, and constraints thoroughly. \\
       - Offer multiple, specific options tailored to unique requirements. \\
       - Anticipate additional needs and provide proactive planning. \\
       - Include relevant examples to support recommendations. \\
       - Consider practical aspects like timing, availability, and potential challenges. \\
       - Provide alternatives for user customization and flexibility. \\
     \\
    4. **Structure and Format Your Solution**  \\
       - Begin with a brief introduction outlining the personalized plan. \\
       - Detail each main component (e.g., accommodation, activities, budget). \\
       - Use markdown format for a visually rich and engaging presentation: \\
         - Utilize headings (\#\#, \#\#\#) and subheadings for clear organization. \\
         - Employ bullet points and numbered lists for easy readability. \\
         - Create tables to present organized information. \\
         - Use bold and italic text for emphasis on key points. \\
         - Incorporate emojis throughout for visual appeal and quick reference. \\
       - Use HTML format if needed for enhanced visual presentation. \\
       - Explicitly reference relevant user need(s) using assigned Need IDs after each major section. \\
       - Ensure the solution is visually appealing and easy to navigate. \\
     \\
    5. **Review and Refine**  \\
       - Verify that all user needs have been addressed. \\
       - Ensure the solution is cohesive, logical, and flows well. \\
       - Check that all Need IDs are correctly referenced. \\
       - Confirm effective use of emojis and rich text formatting throughout. \\
       - Conclude with a summary of key points and invite further questions. \\
     \\
    6. **Finalize and Submit**  \\
       - Save the completed solution using the \texttt{\`}write\_solution\texttt{\`} function. \\
       - Conclude your solution with \texttt{\`}[SolutionEnd]\texttt{\`} to signify completion. \\
     \\
    \#\# Communication Guidelines  \\
     \\
    - Maintain a polite, friendly, and professional tone throughout. \\
    - Provide clear, concise explanations for each aspect of the plan. \\
    - Use engaging language to bring the solution to life and excite the user. \\
    - Tailor communication style to the user's context and request nature. \\
    - Be confident in recommendations while remaining open to adjustments. \\
    - Ensure all explanations and recommendations are user-centric and value-adding. \\
    \\
    \textbf{CONTINUE ON THE NEXT PAGE} \\
    \\
    \\
    \bottomrule
    \end{tabular}
    \caption{The prompt of Solution Craft Agent.}
    \label{tab:solution}
\end{table*}

\begin{table*}[htbp]
    \setlength{\abovecaptionskip}{0.1cm}
    \setlength{\belowcaptionskip}{-0.5cm}
    \centering
    \footnotesize
    \begin{tabular}{p{15cm}}
    \toprule
    \#\# Example \\
    You should write your solution in markdown format and refer to the following example: \\
    \texttt{\`}\texttt{\`}\texttt{\`}markdown \\
    \# Personalized Paris Trip Plan for the Smith Family \\
    Dear Smith Family, \\
     \\
    I'm thrilled to present your customized 6-day Paris itinerary! This plan has been carefully crafted to meet all your specific needs and preferences. Let's explore the exciting details of your Parisian adventure! \\
     \\
    \#\# 1. Accommodation \\
     \\
    **Recommended**: Hotel du Louvre \\
    - **Location**: 1st arrondissement \\
    - **Room Type**: Family Suite (accommodates 4 people comfortably) \\
    - **Key Features**:  \\
      - Central location \\
      - Walking distance to major attractions \\
      - Family-friendly amenities \\
     \\
    > This hotel provides the perfect balance of luxury, comfort, and convenience for your family of four, situated in the heart of Paris. \texttt{\`}(Need ID: 001), (Need ID: 010)\texttt{\`} \\
     \\
    \#\# 2. Transportation  \\
     \\
    **Recommended**: 6-day Paris Visite pass (zones 1-5) \\
    - **Coverage**: All public transportation (metro, RER, buses) \\
    - **Benefits**:  \\
      - Unlimited travel \\
      - Cost-effective for families \\
      - Convenient for exploring different areas of Paris \\
     \\
    > This pass offers flexibility and ease of use, perfect for a family wanting to explore Paris without the hassle of buying individual tickets. \texttt{\`}(Need ID: 002)\texttt{\`} \\
     \\
    \#\# 3. Activities \\
     \\
    \#\#\# 1. Classic Tourist Spots \\
    1. **Eiffel Tower** \\
       - Book skip-the-line tickets in advance \\
       - Best time to visit: Early morning or during sunset \\
    2. **Louvre Museum** \\
       - Consider a guided family tour \\
       - Don't miss: Mona Lisa, Venus de Milo, Winged Victory \\
    3. **Notre-Dame Cathedral** \\
       - Currently closed for renovation \\
       - Admire the exterior architecture \\
     \\
    \#\#\# 2. Child-Friendly Activities \\
    1. **Disneyland Paris** \\
       - Plan for a full day \\
       - Book FastPass tickets to avoid long queues \\
    2. **Jardin d'Acclimatation** \\
       - Amusement park and garden in the Bois de Boulogne \\
       - Perfect for a half-day visit \\
    3. **Cité des Sciences et de l'Industrie** \\
       - Interactive science museum with a children's section \\
       - Planetarium shows available (book in advance) \\
     \\
    > This carefully curated mix of activities ensures that both adults and children in your family will have an enriching and enjoyable experience in Paris. \texttt{\`}(Need ID: 003, Need ID: 004)\texttt{\`} \\
    \\
    \textbf{CONTINUE ON THE NEXT PAGE} \\
    \\
    \\
    \bottomrule
    \end{tabular}
    \caption{The prompt of Solution Craft Agent.}
    \label{}
\end{table*}

\begin{table*}[htbp]
    \setlength{\abovecaptionskip}{0.1cm}
    \setlength{\belowcaptionskip}{-0.5cm}
    \centering
    \footnotesize
    \begin{tabular}{p{15cm}}
    \toprule
    \#\# 4. Dining \\
 \\
    1. **Authentic French Cuisine**: Le Bistrot Paul Bert \\
       - Make a dinner reservation in advance \\
       - Known for classic French dishes and excellent wine list \\
    2. **Variety of Options**: Le Marché des Enfants Rouges \\
       - Oldest covered market in Paris \\
       - Various food stalls offering different cuisines \\
    3. **Child-Friendly Experience**: Café de Flore \\
       - Historic café with a kids' menu \\
       - Famous for its hot chocolate and people-watching opportunities \\
     \\
    > These dining options allow you to experience authentic French gastronomy while ensuring there are suitable and exciting choices for the children. \texttt{\`}(Need ID: 005)\texttt{\`} \\
     \\
    \#\# 5. Budget Breakdown \\
     \\
    Estimated total cost for 6 days (family of four): \$4,000 - \$5,000 \\
    \begin{verbatim}
| Category      | Estimated Cost |
|---------------|----------------|
| Accommodation | $1,800 - $2,200|
| Transportation| $200 - $250    |
| Activities    | $1,000 - $1,300|
| Dining        | $800 - $1,000  |
| Miscellaneous | $200 - $250    |
    \end{verbatim}
    > This budget breakdown balances quality experiences with cost-effective choices, staying within your specified moderate budget range for a family trip to Paris. \texttt{\`}(Need ID: 006)\texttt{\`} \\
     \\
    \#\# 6. Daily Itinerary  \\
     \\
    Here's a day-by-day breakdown of your Paris adventure: \\
     \\
    \#\#\# Day 1: Arrival and Settling In \\
    -  Arrive at Charles de Gaulle Airport \\
    -  Transfer to Hotel du Louvre (consider pre-booking a shuttle) \\
    -  Check-in and freshen up \\
    -  Evening stroll along the Seine River \\
    -  Dinner at Café de Flore \\
     \\
    This daily itinerary balances major attractions, child-friendly activities, and authentic Parisian experiences. It's designed to make the most of your time while allowing for a comfortable pace suitable for a family with children. (Need ID: 007) \\
     \\
    \#\# Final Notes  \\
     \\
    -  Remember to book your activities and restaurants in advance where possible. \texttt{\`}(Need ID: 008)\texttt{\`}
    -  Always carry your Paris Visite pass and a map of the metro system. \texttt{\`}(Need ID: 009)\texttt{\`}
    -  Don't hesitate to ask hotel staff for recommendations or assistance. \texttt{\`}(Need ID: 010)\texttt{\`}
    -  Be flexible with the itinerary – if the children are tired, consider taking a break or swapping activities.
     \\
    I hope this personalized plan exceeds your expectations for your family trip to Paris. If you need any modifications or have any questions, please don't hesitate to ask. Bon voyage!  \\
    \\
    {[SolutionEnd]} \\
    \texttt{\`}\texttt{\`}\texttt{\`} \\
    Remember, your task is not limited to travel planning. Adapt your approach to the specific nature of each user's request, ensuring a comprehensive, personalized, and visually appealing solution that directly addresses their unique needs. \\
     \\
    \#\# Language use \\
    At the beginning of conversation, you should decide the language used to chat with user. \\ 
    - **All of your response must be in English!** \\
    \\
    \\
    \bottomrule
    \end{tabular}
    \caption{The prompt of Solution Craft Agent.}
    \label{}
\end{table*}

\end{document}